\documentclass[superscriptaddress,11pt]{revtex4}

\usepackage{amssymb,amsmath,array,verbatim}
\usepackage[dvipdfmx]{graphicx}

\newcommand{\beq}{\begin{eqnarray}}
\newcommand{\eeq}{\end{eqnarray}}
\newcommand{\p}{\partial}

\newcommand{\vs}[1]{\vspace{#1 mm}}
\newcommand{\hs}[1]{\hspace{#1 mm}}
\newcommand{\bpm}{\begin{pmatrix}}
\newcommand{\epm}{\end{pmatrix}}
\newcommand{\Z}{\mathbb{Z}}
\newcommand{\R}{\mathbb{R}}
\newcommand{\C}{\mathbb{C}}
\newcommand{\tr}{{\rm Tr}}
\newcommand{\D}{\mathcal D}

\newcommand{\ba}{\left(\begin{array}}
\newcommand{\ea}{\end{array} \right)}

\makeatletter
\@addtoreset{equation}{section}

\makeatother

\begin{document}

\title{
Resurgence Structure to All Orders of Multi-bions \\
in Deformed SUSY Quantum Mechanics}

\author{Toshiaki Fujimori}
\email{toshiaki.fujimori018(at)gmail.com}
\address{Department of Physics, and Research and 
Education Center for Natural Sciences, 
Keio University, 4-1-1 Hiyoshi, Yokohama, Kanagawa 223-8521, Japan
}

\author{Syo Kamata}
\email{skamata(at)rikkyo.ac.jp}
\address{Physics Department and Center for Particle and Field Theory, Fudan University,
220 Handan Rd., Yangpu District, Shanghai 200433, China
}
\address{Department of Physics, and Research and 
Education Center for Natural Sciences, 
Keio University, 4-1-1 Hiyoshi, Yokohama, Kanagawa 223-8521, Japan
}

\author{\\ Tatsuhiro Misumi}
\email{misumi(at)phys.akita-u.ac.jp}
\address{Department of Mathematical Science, Akita 
University, 1-1 Tegata Gakuen-machi,
Akita 010-8502, Japan
}
\address{Department of Physics, and Research and 
Education Center for Natural Sciences, 
Keio University, 4-1-1 Hiyoshi, Yokohama, Kanagawa 223-8521, Japan
}
\address{iTHEMS, RIKEN,
2-1 Hirasawa, Wako, Saitama 351-0198, Japan
}

\author{Muneto Nitta}
\email{nitta(at)phys-h.keio.ac.jp}
\address{Department of Physics, and Research and 
Education Center for Natural Sciences, 
Keio University, 4-1-1 Hiyoshi, Yokohama, Kanagawa 223-8521, Japan
}

\author{Norisuke Sakai}
\email{norisuke.sakai(at)gmail.com}
\address{Department of Physics, and Research and 
Education Center for Natural Sciences, 
Keio University, 4-1-1 Hiyoshi, Yokohama, Kanagawa 223-8521, Japan
}
\address{iTHEMS, RIKEN,
2-1 Hirasawa, Wako, Saitama 351-0198, Japan
}

\begin{abstract} \vs{10}
We investigate the resurgence structure 
in quantum mechanical models originating 
in 2d non-linear sigma models 
with emphasis on nearly supersymmetric 
and quasi-exactly solvable parameter regimes. 
By expanding the ground state energy 
in powers of a supersymmetry-breaking 
deformation parameter $\delta \epsilon$, 
we derive exact results for the expansion coefficients. 
In the class of models described by real multiplets, 
the ${\mathcal O}(\delta\epsilon)$ ground state energy has 
a non-Borel summable asymptotic series, 
which gives rise to imaginary ambiguities 
leading to rich resurgence structure. 
We discuss the sine-Gordon quantum mechanics (QM) 
as an example and show that 
the semiclassical contributions from complex multi-bion solutions 
correctly reproduce the corresponding part 
in the exact result including the imaginary ambiguities. 
As a typical model described by chiral multiplets, 
we discuss the $\C P^{N-1}$ QM and show that 
the exact ${\mathcal O}(\delta\epsilon)$ ground state energy 
can be completely reconstructed 
from the semiclassical multi-bion contributions. 
Although the ${\mathcal O}(\delta\epsilon)$ ground state energy 
has trivial resurgence structure, 
a simple but rich resurgence structure appears 
at ${\mathcal O}(\delta\epsilon^{2})$. 
We show the complete cancellation between 
the ${\mathcal O}(\delta\epsilon^{2})$ imaginary ambiguities 
arising from the non-Borel summable perturbation series 
and those in the semiclassical contributions of 
$N-1$ complex bion solutions.
We also discuss the resurgence structure of 
a squashed ${\mathbb C}P^1$ QM.
\end{abstract}

\maketitle

\newpage

\tableofcontents

\newpage 


\section{Introduction}
\label{sec:Intro}
Resurgence theory and trans-series formalism 
\cite{Ec1,Pham1,BH1,Howls1,DH1,Costin1,Sauzin1,Sauzin2} 
in quantum theories 
have shed new light on the nonperturbative analysis 
and the definition of the path integral formalism 
in quantum mechanics (QM)
\cite{Brezin:1977ab, Lipatov:1977cd, Bogomolny:1980ur, ZinnJustin:1981dx, ZinnJustin:1982td, ZinnJustin:1983nr, Alvarez1,Alvarez2,Alvarez3,ZinnJustin:2004ib, ZinnJustin:2004cg, Jentschura:2010zza, Jentschura:2011zza, Dunne:2013ada, Basar:2013eka, Dunne:2014bca,Escobar-Ruiz:2015nsa,Escobar-Ruiz:2015nsa2,Misumi:2015dua,Behtash:2015zha,Behtash:2015loa,Gahramanov:2015yxk,Dunne:2016qix,Fujimori:2016ljw,Sulejmanpasic:2016fwr,Dunne:2016jsr,Kozcaz:2016wvy,Serone:2016qog,Basar:2017hpr,Fujimori:2017oab,Serone:2017nmd,Behtash:2017rqj,Costin:2017ziv,Alvarez:2017sza},
2d quantum field theories (QFT)
\cite{Dunne:2012ae, Dunne:2012zk,Cherman:2013yfa,Cherman:2014ofa,Misumi:2014jua,Misumi:2014bsa,Misumi:2014rsa,Nitta:2014vpa,Nitta:2015tua,Behtash:2015kna,Dunne:2015ywa,Misumi:2016fno,Demulder:2016mja,Sulejmanpasic:2016llc},
4d (or 3d) QFT
\cite{Argyres:2012vv, Argyres:2012ka,Dunne:2015eoa,Dunne:2015eaa,Buividovich:2015oju,Dunne:2016nmc, Gukov:2016njj, Yamazaki:2017ulc},
matrix models and topological string theories
\cite{Marino:2006hs,Marino:2007te, Marino:2008ya, Marino:2008vx, Pasquetti:2009jg,Garoufalidis:2010ya, Drukker:2010nc, Aniceto:2011nu, Marino:2012zq,Schiappa:2013opa,Hatsuda:2013oxa,Aniceto:2013fka,Santamaria:2013rua,Grassi:2014cla,Couso-Santamaria:2014iia,Grassi:2014uua,Couso-Santamaria:2015wga, Aniceto:2015rua,Dorigoni:2015dha,Hatsuda:2015qzx,Franco:2015rnr,Couso-Santamaria:2016vcc,Kuroki:2016ucm,Couso-Santamaria:2016vwq,Arutyunov:2016etw},
and localization-applicable supersymmetric Yang-Mills (SYM) theories
\cite{Aniceto:2014hoa, Honda:2016mvg, Honda:2016vmv}.
They have also been discussed 
in terms of the exact WKB analysis of 
Schr\"odinger-type ordinary differential equations 
\cite{Balian:1978ab, Voros1, DDP1, CNP1, DLS1, DP1, Takei1, CDK1, Takei2, Getm1, AKT1, Schafke1, Getm2}.
In the resurgence theory, 
the Borel resummations of perturbation series 
around all nontrivial backgrounds are taken into account, 
and it is expected that such a full semiclassical expansion 
(resurgent trans-series) leads to 
an unambiguous definition of quantum theories \cite{K1,K2,K3}.

In the original argument of the resurgent expansion
\cite{Bogomolny:1980ur,ZinnJustin:1981dx,ZinnJustin:1982td,ZinnJustin:1983nr,Alvarez1,Alvarez2,Alvarez3,ZinnJustin:2004ib,ZinnJustin:2004cg, Jentschura:2010zza,Jentschura:2011zza}, 
one needs to take account of configurations 
composed of instanton-antiinstanton pairs called ``bions"
\cite{Unsal:2007vu,Unsal:2007jx,Shifman:2008ja,Poppitz:2009uq,Anber:2011de,Poppitz:2012sw,Misumi:2014raa}. 
Imaginary ambiguities emerging from 
such bion contributions cancel out 
those arising in the non-Borel-summable perturbation series.
Recent studies have manifested 
the true nature of the bion configurations 
from the viewpoint of the complexified path integral,
where each bion configuration emerges 
as a complex saddle point \cite{Behtash:2015zha, Behtash:2015loa,Fujimori:2016ljw,Fujimori:2017oab}. 
In the framework of the complexified theory, 
the original integration contour of the path integral 
is decomposed into the so-called ``Lefschetz thimbles"
\cite{Witten:2010cx,Cristoforetti:2013wha,Fujii:2013sra,Tanizaki:2014tua,Tanizaki:2014xba,Kanazawa:2014qma,Tanizaki:2015tnk,DiRenzo:2015foa,Fukushima:2015qza,Tanizaki:2015rda,Fujii:2015bua,Alexandru:2016gsd,Tanizaki:2016xcu}, 
each of which is associated with one of the saddle points. 
The contribution from each bion background 
in the resurgent trans-series 
is given by the path integral along the associated Lefschetz thimble. 
Those deformed contours vary depending on 
the complexified coupling constant (${\rm arg} \, g^{2} \not = 0$)
and some of the integration cycles 
discontinuously jump at ${\rm arg} \, g^{2}=0$. 
Such a discontinuity of a thimble gives rise to 
an imaginary ambiguity reflecting ``the Stokes phenomenon" 
in the corresponding sector of the trans-series. 
Those ambiguities are expected to cancel among themselves 
and hence there is no ambiguity in the entire trans-series, 
which corresponds to the path integral along the original contour. 

In \cite{Behtash:2015zha, Behtash:2015loa} 
and our previous work \cite{Fujimori:2016ljw, Fujimori:2017oab}, 
exact solutions of the holomorphic equations of motion 
were investigated in the double-well, sine-Gordon and 
${\mathbb C}P^{1}$ QM with fermionic degrees of freedom (incorporated as a parameter $\epsilon$), 
and it was shown that the bions appear 
as the exact solutions of the complexified equation of motions.
In \cite{Fujimori:2017oab}, 
an infinite tower of exact multi-bion solutions were found, 
and the exact resurgent trans-series was obtained 
to all orders in the perturbative and nonperturbative expansion. 
It was shown that the response of the exact ground state energy 
under a deformation from the supersymmetric (SUSY) point 
($\epsilon=1$) can be expressed as 
a trans-series with nonperturbative terms 
corresponding to multi-bion saddle points, 
together with perturbation series around them.
In the trans-series, the imaginary ambiguity 
associated with the non-Borel summable perturbation series 
around the $p$-bion background is canceled 
by that arising from the semiclassical contribution of 
$(p+1)$-bion saddle point.  
By applying the Lefschetz thimble method, 
all the semiclassical contributions from the multi-bion solutions 
are shown to agree with the corresponding parts in the exact result.

In this work, we investigate resurgence structure 
in a broader class of quantum mechanical models
from the viewpoint of the complexified path integral 
and its complex multi-bion saddle points.
We focus on SUSY and quasi-exactly solvable (QES) QM, 
where we can take advantage of exact results 
to probe the resurgence structure of those models.
By introducing a SUSY-breaking (or QES-breaking) 
deformation parameter $\delta\epsilon$
and expanding the ground state energy 
around the SUSY (or QES) point in powers of $\delta\epsilon$, 
we reveal all order multi-bion contributions 
with nontrivial resurgence structure. 

We classify the models into two classes: 
(i) quantum mechanics on a Riemannian manifold 
described by real multiplets and 
(ii) quantum mechanics on a K\"ahler manifold 
described by chiral multiplets. 
In both classes, the ground state energy 
at the SUSY and QES points $(\delta \epsilon = 0)$ 
does not receive any nonperturbative correction 
due to cancellation among various 
(real and complex) multi-bion contributions.
In the first class, 
the $\mathcal O(\delta \epsilon)$ ground state energy
has a non-Borel-summable asymptotic series 
in each sector of the trans-series, 
which gives rise to an imaginary ambiguity 
leading to rich resurgence structure.
In the second class, a localization method \cite{Atiyah-Bott, DT} 
is (partially) available to determine 
the $\mathcal O(\delta \epsilon)$ ground state energy, 
which leads to simpler (sometimes trivial) resurgence structure 
than that in the first class. 

As typical examples in the two classes, 
we consider the sine-Gordon QM (the first class) and 
the (squashed) ${\mathbb C}P^{N-1}$ QM (the second class).
In the sine-Gordon QM, 
we obtain the exact result 
for the ${\mathcal O}(\delta\epsilon)$ ground state energy,
which is composed of a perturbation series and 
an infinite tower of nonperturbative terms, 
each of which has a non-Borel-summable asymptotic series. 
Based on the complexified path integral,
we show that the semiclassical multi-bion contributions 
reproduce the corresponding parts 
in the exact result including the imaginary ambiguities 
which cancel those in the other sectors.
This supports the resurgence to all orders 
in the nonperturbative exponential. 
In the ${\mathbb C}P^{N-1}$ QM,
we find $(N-1)$ types of (real and complex) bion solutions and 
show that the exact result 
for the ${\mathcal O}(\delta\epsilon)$ ground state energy 
can be completely reconstructed 
from the semiclassical multi-bion contributions. 
We determine the non-Borel summable perturbation series 
of the ${\mathcal O}(\delta\epsilon^2)$ ground state energy
to all orders in $g^2$ and confirm that 
its imaginary ambiguities are canceled 
by those in the single-bion contributions.
We also show by deforming the target space that 
nontrivial resurgence structure can be seen 
in the ${\mathcal O}(\delta\epsilon)$ ground state energy 
for a generic target manifold.

The organization of this paper is as follows. 
In Sec.\,\ref{sec:SQM}, 
we discuss generic properties of the ground state energy 
in the two classes of SUSY models. 
By expanding in terms of 
a SUSY-breaking parameter $\delta \epsilon$, 
we show that ``the generating function" $\langle 0 | 0 \rangle$ 
plays an important role 
in determining the resurgence structure of the ground state energy 
around the SUSY point in the parameter space. 
In Sec.\,\ref{sec:SG}, we investigate the resurgence structure 
in the sine-Gordon QM.
The $\mathcal O (\delta \epsilon)$ ground state energy is 
derived exactly and compared with 
the semiclassical multi-bion contributions. 
In Sec.\,\ref{sec:CPN}, we investigate the resurgence structure  
in the ${\mathbb C}P^{N-1}$ QM, 
with emphasis on the exact complex solutions 
and their contributions to the ground state energy 
around the SUSY and QES points.
In Sec.\,\ref{sec:sCPN}, we investigate the resurgence structure 
in the squashed ${\mathbb C}P^{1}$ QM.
Sec.\,\ref{sec:SD} is devoted to summary and discussion. 
Appendices \ref{appendix:SQM}, \ref{app:Loc}, \ref{appendix:E2CPN}, 
\ref{appendix:quasi_moduli}, \ref{appendix:effpot}, \ref{appendix:QMI}, 
and \ref{appendix:one-loop} are devoted to
the supersymmetric QM, the localization method, 
the perturbative part in the ${\mathbb C}P^{N-1}$ QM,
the quasi moduli space, the kink-antikink effective potential, 
the Lefschetz thimble analysis, and the one-loop determinant, respectively.


\section{SUSY Quantum Mechanics with Deformation Parameter}
\label{sec:SQM}

\subsection{Quantum Mechanics on Riemannian manifolds}

In this paper, we discuss quantum mechanics of a particle on 
a manifold $\mathcal M$ with a potential $V$. 
The Schr\"odinger equation takes the form
\beq
H \Psi = \left[ -g^{2}\Delta + V \right] \Psi\,, \hs{10} 
\Delta \Psi \equiv \frac{1}{\sqrt{\det G}} \partial_{i} \left( G^{ij} \sqrt{\det G} \right) \partial_{j} \Psi,
\label{eq:schrodinger}
\eeq
where $G_{ij}$ is the metric of the target manifold $\mathcal M$,
$\Delta$ is the Laplacian, 
and $\partial_{i} \equiv {\partial\over{\partial \varphi^{i}}}$ 
stand for the partial derivatives with respect to 
the real bosonic variables 
$\varphi^{i}$ ($i=1,..,n={\rm dim} \, \mathcal M$)
corresponding to the coordinates on $\mathcal M$. 
The focus of this paper is on the class of models 
which can be obtained 
from supersymmetric quantum mechanical models 
by a SUSY-breaking deformation. 
In particular, we put a special emphasis on 
the expansion around SUSY and quasi-exactly solvable (QES) points. 
In such a model, the Hamiltonian projected onto 
the lowest fermion number eigenspace
takes the form of \eqref{eq:schrodinger} 
with a bosonic potential of the form 
(see Appendix \ref{appendix:SQM} for details)
\beq
V = \frac{1}{g^2} G^{ij} \partial_{i} W \partial_{j} W - \epsilon \Delta W, 
\label{eq:H_SQM}
\eeq
where $W$ is a real function on $\mathcal M$, which we call the superpotential, 
and $\epsilon$ is the SUSY-breaking deformation parameter. 

The model with $\epsilon=1$ corresponds to the SUSY case, 
where the exact wave function for the SUSY ground state 
$|0\rangle$ is obtained as 
\beq
\Psi^{(0)} \equiv \langle \varphi | 0\rangle  = \exp \left( -{W\over{g^{2}}} \right) \,.
\label{eq:WF_SQM}
\eeq
This SUSY invariant state is well-defined only when it is normalizable
\beq
\langle 0 | 0 \rangle = \int_{\mathcal M} dv \exp \left( - \frac{2W}{g^2} \right) < \infty \,,
\label{eq:00_SQM}
\eeq
where $dv$ is the volume form on $\mathcal M$: 
$dv = \sqrt{\det G} \, d \varphi^1 \wedge \cdots \wedge d \varphi^n$.
For example, in a single-variable case, 
a polynomial superpotential $W(\varphi)$ of even degree 
satisfies the normalizability condition
while that of odd degree does not.
This means that the present setup includes 
triple-well, 5-well,..., $(2n+1)$-well potentials
while it does not include double-well, 4-well,..., $2n$-well potentials.
As is well known, SUSY is spontaneously broken 
in the latter cases. 
 
We will investigate the ground state energy and the wave function 
by expanding them in powers of the SUSY-deformation parameter 
$\delta \epsilon = \epsilon -1$ as $E = \delta \epsilon E^{(1)} + \delta \epsilon^2 E^{(2)} + \cdots$
and $\Psi = \Psi^{(0)} + \delta \epsilon \Psi^{(1)} + \cdots$.
In other words, we consider their responses 
under the SUSY-breaking deformation
\beq
E^{(n)} = \frac{1}{n!} \frac{\p^n}{\p \epsilon^n} E \bigg|_{\epsilon=1}, \hs{10}
\Psi^{(n)} = \frac{1}{n!} \frac{\p^n}{\p \epsilon^n} \Psi \bigg|_{\epsilon=1}. 
\eeq
These expansion coefficients can be determined 
by the standard Rayleigh-Schr\"odinger perturbation theory
\beq
E^{(1)} = \frac{\langle 0| \delta H |0\rangle }{\langle 0 | 0 \rangle}\,, \hs{10}
E^{(2)} = - \frac{\langle \Psi^{(1)}| H_{\epsilon=1} |\Psi^{(1)}\rangle}{\langle 0 | 0 \rangle}, \hs{10} 
\cdots \,,
\label{eq:E1E2_SQM}
\eeq
where $\delta H = -\Delta W$ is 
the SUSY-breaking perturbation Hamiltonian. 

For the purpose of understanding the resurgence structure 
at each order of $\delta \epsilon$, 
the property of the denominator $\langle 0 |0\rangle$ 
in Eq.\,\eqref{eq:E1E2_SQM} is of great importance. 
Since this gives the vacuum expectation value (VEV) of $W$ 
when differentiated with respect to $1/g^2$, 
we call this quantity the generating function of $W$ in the following. 
Applying the saddle point method, 
we obtain the following schematic form of the generating function
\beq
\langle 0 |0 \rangle \ = \ \int_{\mathcal M} dv \exp \left( -{2 W \over{g^{2}}}\right) 
\ = \ \sum_{s \in \mathfrak S} F_s(g^2) \exp \left( -{2 W_s \over{g^{2}}} \right), 
\eeq
where $\mathfrak S$ denotes the set of the saddle points of $W$, 
$W_s$ is the value of $W$ at the saddle point $s$ and 
$F_s(g^2)$ is the perturbation series around the saddle point $s$. 

Depending on the type of SUSY models 
from which we obtain the bosonic potential, 
the generating function $\langle 0 |0\rangle$ has different properties. 
If the target space $\mathcal M$ is a Riemannian manifold  
whose coordinates are parametrized 
by the bosonic components of ``real multiplets", 
the functions $F_s(g^2)$ have nontrivial asymptotic expansions, 
which give complicated resurgent structure to $E^{(n)}$. 
This class includes the sine-Gordon QM and 
the $(2n+1)$-well models ($n \in {\mathbb Z}$).
While if the target space is a K\"ahler manifold 
whose complex coordinates constitute ``chiral multiplets", 
the functions $F_s(g^2)$ are monomials of $g^2$, 
which leads to a simplified 
(sometimes trivial) resurgent trans-series in $E^{(n)}$.
Examples of this type are the $\C P^{N-1}$ QM, 
the Grassmannian QM and their deformed models.

\subsection{Real multiplet: generating function and Lefschetz thimble}

Let us first see some generic properties of 
the generating function $\langle 0 | 0 \rangle$ 
in the SUSY QM described by real multiplets. 
After the projection onto the lowest fermion number states,
the action takes the form 
(see Appendix \ref{appendix:SQM} for details)
\beq
S = \int d t \left( \frac{1}{4g^2} G_{ij} \dot \varphi^i \dot \varphi^j 
- \frac{1}{g^2} G^{ij} \p_i W \p_j W + \Delta W \right).
\eeq
This class of models includes the SUSY ($2n+1$)-well model 
and the SUSY sine-Gordon model.

As we have discussed, 
the generating function for $W$ can be obtained 
by evaluating the integral 
\beq
\langle 0 |0 \rangle = \int_{\mathcal M} dv \exp \left( -{2 W \over{g^{2}}}\right) \,. 
\eeq
Let us apply the Lefschetz thimble method to this integral. 

We assume that the target manifold $\mathcal M$ 
has a suitable complexification $\mathcal M^\C$
parametrized by the complexified coordinates $\varphi^i$
and the function $W$ can be analytically continued 
to $\mathcal M^\C$ as a holomorphic function of $\varphi^i$. 
The thimble $\mathcal J_s$ associated with a saddle point $s$ of $W$ 
is a middle dimensional subspace of $\mathcal M^\C$ 
which can be reached by the upward flows from the saddle point $s$: 
\beq
\mathcal G_{i \bar j} \frac{d \varphi^i}{dt} = \overline{\frac{1}{g^2} \frac{\p W}{\p \varphi^j}}, \hs{10}
\lim_{t \rightarrow -\infty} \varphi^i = \varphi^i_s, 
\label{eq:flow}
\eeq
where $t$ is a formal flow parameter and 
$\mathcal G_{i \bar j}$ is a suitable positive definite metric 
on $\mathcal M^\C$ such that $\mathcal G_{i \bar j} = G_{ij}$ 
on $\mathcal M$. 
The dual thimble $\mathcal K_s$ is defined as 
a middle dimensional subspace of $\mathcal M^\C$ 
\textbf{(${\rm dim}\,{\mathcal J}_s = {\rm dim}\,{\mathcal K}_s$)}
which flows to the saddle point $s$: 
\beq
\mathcal G_{i \bar j} \frac{d \varphi^i}{dt} = \overline{\frac{1}{g^2} \frac{\p W}{\p \varphi^j}}, \hs{10}
\lim_{t \rightarrow \infty} \varphi^i = \varphi^i_s. 
\eeq
The definition of the thimble $\mathcal J_s$ 
and its dual $\mathcal K_s$ implies that 
the real and imaginary parts of the complexified superpotential satisfy
\beq
{\rm Re} \, W |_{\mathcal J_s} \geq {\rm Re} \, W_s \geq {\rm Re} \, W |_{\mathcal K_s}, \hs{10}
{\rm Im} \, W |_{\mathcal J_s} = {\rm Im} \, W_s = {\rm Im} \, W |_{\mathcal K_s}.
\eeq 
It follows from these properties that 
$\mathcal J_s$ and $\mathcal K_s$ can intersect only 
at the saddle point $s$. 
Since the imaginary parts are generically different 
(${\rm Im} \, W_s \not = {\rm Im} \, W_{s'}$) 
at two different saddle points $s \not = s'$ 
for a generic function $W$, 
there is no intersection between 
$\mathcal J_s$ and $\mathcal K_{s'}~(s' \not = s)$.
Provided Stokes phenomena do not arise, 
we can define the intersection pairing of thimbles and their duals as 
\beq
\langle \mathcal J_s, \mathcal K_{s'} \rangle = \delta_{s s'}. 
\eeq
By using this pairing, the original integration cycle 
$\mathcal M$ can be decomposed as  
\beq
\mathcal M = \sum_{s \in \mathfrak S} n_s \mathcal J_s, 
\hs{10} n_s = \langle \mathcal M , \mathcal K_s \rangle \,,
\eeq
where $n_s$ are the intersection numbers 
between $\mathcal M$ and the dual thimble $\mathcal K_s$.
Correspondingly, the generating function can be decomposed as
\beq
\langle 0 |0 \rangle = \sum_{s \in \mathfrak S} F_s(g^2) 
\exp \left( -{2 W_s \over{g^{2}}} \right) \,, 
\eeq
with
\beq
F_s(g^2) = n_s \int_{\mathcal J_s} dv \, \exp 
\left[ - \frac{2(W-W_s)}{g^2} \right]. 
\eeq
However, this decomposition becomes ambiguous 
if there exists a kink solution described by 
the BPS equation in the original SUSY model
\beq
G_{ij} \frac{\p \varphi^j}{\p \tau} = \frac{\p W}{\p \varphi^i},
\eeq
where $\tau$ is interpreted as the Euclidean time. 
Since the BPS kink solution is 
a flow connecting two different saddle points $(s \not = s')$, 
its existence implies that 
$\mathcal J_s$ coincides with $\mathcal K_{s'}$ 
and hence the intersection pairing is ill-defined. 
We can make it well-defined by giving a small imaginary part to $g^2$ as a regularization parameter. 
Although such a complexified coupling constant gives a well-defined intersection pairing, 
it can give different decompositions of $\mathcal M$ 
depending on the sign of $\arg \, g^2=0$. 
This is because the thimbles can have discontinuity at $\arg \, g^2=0$
\beq
\mathcal J_s^+ \not = \mathcal J_s^-, \hs{10} 
\mathcal K_s^+ \not = \mathcal K_s^- \,,
\eeq
where thimbles with $\pm$ denote those 
for positive and negative $\arg \, g^2$. 
Therefore, the generating function can also have different 
decompositions depending on $\arg \, g^2$: 
\beq
\langle 0 |0 \rangle = \sum_{s \in \mathfrak S} F_s^\pm(g^2) \exp \left( -{2 W_s \over{g^{2}}} \right) \,, 
\eeq
with
\beq
F_s^\pm(g^2) = n_s^\pm \int_{\mathcal J_s^\pm} dv \, \exp \left[ - \frac{2(W-W_s)}{g^2} \right], \hs{10}
n_s^\pm \equiv \langle \mathcal M , \mathcal K_s^\pm \rangle \,.
\eeq
This means that a Stokes phenomenon occurs 
on the line ${\rm arg}\,g^{2}=0$,
and consequently the asymptotic series of $\langle 0 | 0 \rangle$ 
at each saddle point has an ambiguity for real $g$.
As we will see below, the Stokes phenomenon 
for the generating function $\langle 0|0\rangle$
gives a nontrivial resurgence structure 
to $E^{(1)} = \langle 0| \delta H|0 \rangle / \langle 0|0\rangle$. 

It is worth noting that the generating function can be written as
\beq
\langle 0 |0 \rangle = e^{-\frac{2W_0}{g^2}} \sum_{s \in \mathfrak S} F_s^\pm(g^2) \exp \left( -2 S_{{\rm kink},s}  \right) \,, 
\label{eq:NP_bion}
\eeq
where $W_0 \equiv {\rm min}_{s \in \mathfrak S} \, W_s$ is 
the value of $W$ at the global minimum of the potential and 
$S_{{\rm kink},s} \equiv (W_s-W_0)/g^2$ are the on-shell values of the Euclidean action for the BPS kink solutions. 
Eq.\,\eqref{eq:NP_bion} implies that pairs of kinks, 
i.e. bions give the nonperturbative contribution to $\langle 0 | 0 \rangle$.


\subsection{Chiral multiplet: generating function and Lefschetz thimble}

Next let us consider the case of a K\"ahler target manifold 
parametrized by the bosonic components of chiral multiplets,
which includes the $\C P^{N-1}$ QM
and the Grassmannian QM models. 
In the previous subsection, 
we have seen that the Stokes phenomenon for $\langle 0 | 0 \rangle$ 
can give nontrivial resurgence structure of $E^{(1)}$. 
Here we see that in the case of a K\"ahler target manifold,
no Stokes phenomenon occurs for $\langle 0 | 0 \rangle$
and hence the expansion coefficients $E^{(n)}$ have 
relatively simple resurgence structure. 

The K\"ahler metric can be written in terms of 
a K\"ahler potential $K$ on each coordinate patch as
\beq
G_{i \bar j} = \p_i \bar \p_{\bar j} K\,,
\label{eq:Kmetric}
\eeq 
where $i, \bar j = 1, \cdots, n$.
For example, the model with $\mathcal M \cong \C P^{N-1}$ is 
given by $K = \log (1 + \varphi^i\bar\varphi^{\bar i})$ 
with the inhomegeneous coordinates $\varphi^i~(i=1,\cdots N-1)$. 
We also that there is a holomorphic isometry,
whose holomorphic Killing vector 
$\Xi = \xi^i \p_i + \bar \xi^{\bar i} \bar \p_{\bar i}$ satisfies
\beq
\p_i \xi_{\bar j} + \bar \p_{\bar j} \bar \xi_i \, = \, 
\bar \p_{\bar j} \xi^i \, = \, 
\p_i \bar \xi^{\bar j} = 0.
\label{eq:KillingEq}
\eeq
We here define the moment map $\mu$ 
for the holomorphic isometry as $d\mu = i_{\Xi} \omega$,
where $\omega$ is the K\"ahler form 
$\omega = i G_{i \bar j} d \varphi^i \wedge d \bar \varphi^{\bar j}$ 
and $i_{\Xi}$ denotes the interior product (contraction) 
with respect to the Killing vector $\Xi$.
In terms of the components, 
the equation for $\mu$ can be rewritten as
\beq
\p_i \mu = - i G_{i \bar j} \bar \xi^{\bar j}, \hs{10} 
\bar \p_{\bar i} \mu = i G_{j \bar i} \xi^j.
\label{eq:momap}
\eeq

The SUSY QM of chiral multiplets we discuss in this paper 
can be obtained from 
the 2d $\mathcal N =(2,0)$ non-linear sigma model 
\beq
S_{2d} = \frac{1}{g_{2d}^2} \int d^2 x \, G_{i \bar j} 
\Big[ - \p_\mu \varphi^i \p^\mu \bar \varphi^{\bar j} 
+ i \bar \psi^{\bar j} (\D_t - \D_x) \psi^i \Big].
\label{eq:S_2d}
\eeq
By imposing the periodic boundary condition twisted by the isometry 
and reducing the spatial direction, 
we obtain the following 1d action
\beq
S = \frac{1}{g^2} \int dt \, G_{i \bar j} 
\left[ \dot \varphi^i \dot{\bar \varphi}^{\bar j} 
- \xi^i \bar \xi^{\bar j} + i \bar \psi^{\bar j} \D_t \psi^i 
- i \nabla_k \xi^i \bar \psi^{\bar j} \psi^k \right],
\label{eq:KahlerL1}
\eeq
where $\D_i \psi^i = \p_i \psi^i + \Gamma^i_{jk} \p_i \varphi^j \psi^k$, 
$\nabla_k \xi^i = \p_k \xi^i + \Gamma^i_{jk} \xi^j$,
$\Gamma_{jk}^i = \p_j G_{k \bar l} G^{\bar l i}$ and 
\beq
\frac{1}{g^2} = \frac{1}{g_{2d}^2} \times 2\pi 
\{\mbox{compactification radius}\}.
\eeq 
Under the compactification with the twisted boundary condition, 
sigma model instantons in the Euclidean 2d non-linear sigma model 
\cite{Polyakov:1975yp} decompose into a set of 
``fractional instantons" \cite{Eto:2004rz,Eto:2006mz,Eto:2006pg} 
(see also \cite{Bruckmann:2007zh,Brendel:2009mp}). 
From the viewpoint of quantum mechanics, 
such fractional instantons appear as the BPS kinks 
(see, e.g., \cite{Dvali:1996bg,Chibisov:1997rc,Isozumi:2004jc,Isozumi:2004va})
carrying a fractional instanton number
\beq
S_{E,\,{\rm inst}} = \frac{1}{g_{2d}^2} \int i G_{i \bar j} \, 
d \varphi^i \! \wedge d \bar \varphi^{\bar j} ~~~ \rightarrow ~~~ 
S_{E,\,{\rm kink}} = \frac{1}{g^2} \int ( \p_i \mu \, d \varphi^i 
+ \bar \p_{\bar i} \mu \, d \bar \varphi^{\bar i}). 
\eeq
Thus it could be speculated that 
there is a close relationship between 
the 2d and 1d nonperturbative effects induced by those objects. 

As discussed in Appendix \ref{appendix:SQM}, 
the projection onto the lowest fermion number eigenstates
gives a bosonic potential of the form \eqref{eq:H_SQM} 
with the identification 
\beq
W \, = \,\mu\,.
\eeq
Thus the Schr\"odinger equation takes form
\beq
H \Psi = G^{\bar j i} \left[- g^2 \p_i \bar \p_{\bar j} 
+ \frac{1}{g^2} \p_i \mu \bar \p_{\bar j} \mu 
- \epsilon \p_i \bar \p_{\bar j} \mu \right] \Psi.
\eeq
As in the case of real multiplets, 
we can find the exact SUSY ground state wave function as
\beq
\Psi^{(0)} \equiv \langle \varphi | 0 \rangle = \exp \left( - \frac{\mu}{g^2} \right). 
\eeq
The normalization factor $\langle 0 | 0 \rangle$ can be regarded as 
the generating function of the vacuum expectation value of $\mu$
\beq
\langle 0 | 0 \rangle = \int_{\mathcal M} dv \, \exp \left( - \frac{2\mu}{g^2} \right), \hs{10} dv = \frac{\omega^n}{n!}. 
\eeq
Again, we can use the saddle point method 
to decompose the generating function
\beq
\langle 0 | 0 \rangle \,=\, \sum_{s \in \mathfrak S} F_s(g^2) \exp \left( - \frac{2\mu_s}{g^2} \right) 
\,=\, e^{-\frac{2\mu_0}{g^2}} \sum_{s \in \mathfrak S} F_s(g^2) \exp \left( -2 S_{{\rm kink},s}  \right) \,,
\label{eq:Kahler_GF}
\eeq
where $\mu_0 \equiv {\rm min}_{s \in \mathfrak S} \mu_s$ is 
the value of the moment map at the global minimum of the potential 
and $S_{{\rm kink},s}$ is the on-shell value of the action 
for the BPS kink satisfying
\beq
G_{i \bar j} \frac{d \varphi^i}{d\tau} = \frac{\p \mu}{\p \bar \varphi^{\bar j}}, \hs{10} 
\lim_{\tau \rightarrow - \infty} \varphi^i = \varphi_0^i, \hs{5} 
\lim_{\tau \rightarrow \infty} \varphi^i = \varphi_s^i.
\eeq
Eq.\,\eqref{eq:Kahler_GF} implies that 
the nonperturbative contributions to $\langle 0 | 0 \rangle $, 
which are proportional to $\exp \left( -2 S_{{\rm kink},s}  \right)$, 
can be regarded as bion contributions. 
The most important property of the generating function 
in the K\"ahler case is that $F_s(g^2)$ can be computed 
by the Duistermaat-Heckman localization formula \cite{Atiyah-Bott, DT} 
\beq
F_s(g^2) = \left( \frac{\pi g^2}{2} \right)^n \frac{1}{\det M_s},
\eeq
where $M_s$ is the $n$-by-$n$ matrix which represents 
the action of the Killing vector $-i \xi$ 
on the tangent space at the saddle point $s$ 
(see Appendix \ref{app:Loc} for details). 
Since $F_s(g^2)$ is not a divergent asymptotic series, 
there is no ambiguity and hence 
the decomposition \eqref{eq:Kahler_GF} is unambiguous. 
This property ensures that 
the expansion coefficients of the ground state energy
\eqref{eq:E1E2_SQM} have relatively simple resurgence structure. 
In particular, when the perturbation Hamiltonian 
is a polynomial of $\mu$: 
\beq
\delta H = -\Delta \mu =  P(\mu) \,,
\label{eq:Lap_momap}
\eeq
the first expansion coefficient $E^{(1)}$ can be calculated 
from the generating function 
$\langle 0|0\rangle =  \int dv \, e^{-\frac{2\mu}{g^2}}$ as
\beq
E^{(1)} \, = \, - \frac{\langle 0| \Delta \mu |0\rangle}{\langle 0 | 0 \rangle} 
\, = \, \frac{P( \hat \mu ) \langle 0 | 0 \rangle}{\langle 0 | 0 \rangle} \,, 
\hs{10} 
\hat \mu \equiv \frac{1}{2} g^4 \frac{\p}{\p g^2} \,.
\label{eq:0Dm0}
\eeq
This implies that $E^{(1)}$ has 
a finite-order $g^{2}$ asymptotic series in each bion sector 
and hence there is no ambiguity.  
This indicates that the resurgence structure is trivial 
at ${\mathcal O}(\delta\epsilon)$ in this type of models.
We will see these properties in detail in the $\C P^{N-1}$ QM 
in Sec.\,\ref{sec:CPN}.

On the other hand, when $P(\mu)$ is not a polynomial of $\mu$, 
$g^{2}$ asymptotic series can be infinite-order (non-Borel-summable) 
and gives an ambiguous contribution to $E^{(1)}$
at the Stokes line ${\rm arg}\,g^{2}=0$. 
Therefore, the resurgence structure can be nontrivial 
at ${\mathcal O}(\delta\epsilon)$. 
We note that it is still relatively simple 
since there is no Stokes phenomenon 
in the denominator $\langle 0 | 0 \rangle$, 
so that imaginary ambiguities cancel 
between adjacent nonperturbative sectors. 
We will see details of these properties 
in the squashed $\C P^{N-1}$ QM in Sec.\,\ref{sec:sCPN}.

It is notable that $E^{(2)}$ has richer resurgence structure 
in any type model with a K\"ahler target manifold.
Nevertheless, since the denominator 
$\langle 0|0\rangle$ is free from ambiguity, 
the resurgence structure is relatively ``clean", 
where imaginary ambiguities arising from perturbation series 
around $p$-bion backgrounds are completely canceled 
by those in semiclassical $(p+1)$-bion contributions. 
We will see these properties 
in the $\C P^{N-1}$ QM in detail in Sec.\,\ref{sec:CPN}.


\section{Resurgence Structure in the Sine-Gordon Quantum Mechanics}
\label{sec:SG}

In this section, we consider the sine-Gordon QM 
as an example of a model described by real multiplets 
discussed in Sec.\,II.
We first derive some exact results 
for the expansion coefficients of the ground state energy 
around the SUSY and QES solvable points. 
Then we look into the multi-bion solution and show 
by applying the Lefschetz thimble method to the quasi-moduli integral
that the semiclassical bion contributions 
with imaginary ambiguities reproduce 
the corresponding parts in the exact results.

\subsection{Sine-Gordon quantum mechanics}
Throughout this section, 
we use $\theta$ instead of $\varphi$ 
as the periodic coordinate of $S^1$. 
The superpotential for the sine-Gordon QM is given by
\beq
W(\theta) = - {m\over{2}} \cos \theta \,.
\label{eq:superPot_SG}
\eeq
After the projection onto the lowest fermion number eigenspace, 
the Hamiltonian with the SUSY-breaking deformation takes the form
\beq
H \Psi = \left[ -g^2 \p_\theta^2 + \frac{m^2}{4g^2} \sin^2 \theta - \frac{m\epsilon}{2} \cos \theta \right] \Psi. 
\label{eq:Schroedinger_eq}
\eeq
This sine-Gordon model becomes supersymmetric at $\epsilon=1$. 

Since the potential of the sine-Gordon QM is a periodic function, 
its energy spectrum has a band structure. 
Energy eigenstates within each band are characterized 
by the Bloch angle $2\pi\alpha$ 
defined as a twisted angle of boundary conditions 
\beq
\Psi(\theta+2\pi) = e^{2 \pi i \alpha} \Psi(\theta).  
\label{eq:bloch_SG}
\eeq
The Bloch angle takes values between $0\le \alpha <1$ 
since $\alpha$ and $\alpha+{\mathbb Z}$ are equivalent. 
The factor $2\pi \alpha$ can be eliminated 
from the boundary condition by the redefinition 
$\Psi(\theta) \rightarrow \tilde \Psi(\theta)\equiv e^{i \alpha \theta} \Psi(\theta)$. 
Then the Sch\"odinger equation \eqref{eq:Schroedinger_eq} 
for the periodic wave function $\tilde \Psi$ becomes 
\beq
H \tilde \Psi &=& 
\left[ -g^2 (\p_\theta - i \alpha)^2 + \frac{m^2}{4g^2} \sin^2 \theta - \frac{m\epsilon}{2} \cos \theta \right] \tilde \Psi 
~=~ E \tilde \Psi.
\label{eq:Schoe_SG2}
\eeq
In the following, we discuss the ground state of the system 
described by this Sch\"odinger equation. 
The ground state energy can be read off from 
the low temperature limit ($\beta \rightarrow \infty$) 
of the partition function, 
which can be defined for each $\alpha$ 
by the following Euclidean path integral 
over periodic configuration 
$\theta(\tau+\beta) = \theta(\tau) \ {\rm mod} \, 2\pi$
\beq
Z(\alpha) = \tr [ e^{-\beta H} ] = \int \D \theta \, \exp (- S_E), 
\label{eq:Zalpha_SG}
\eeq
where the Euclidean action is given by
\beq
S_E = \int_0^\beta d\tau \left[ \frac{1}{4g^2} \left( \dot \theta^2 + m^2 \sin^2 \theta \right) 
- \frac{m\epsilon}{2} \cos \theta - i \alpha \dot \theta \right].
\label{eq:Salpha_SG}
\eeq
The last term is the topological term related to the Bloch angle,
which gives $\mp \pi i \alpha$ for each (anti-)kink
corresponding to the tunneling process 
between the local and global minima of the potential 
($\theta=\pi$ and $\theta=0$).
It is notable that the Hamiltonian can be rewritten as
\beq
H = \bar Q Q - \frac{m}{2} \delta \epsilon \cos \theta,
\label{eq:HQQ_SG}
\eeq
where $\delta \epsilon =\epsilon - 1$ and
\beq
Q = i g \left( \p_\theta - i \alpha + \frac{m}{2g^2} \sin \theta \right), \hs{10} 
\bar Q = i g \left( \p_\theta - i \alpha - \frac{m}{2g^2} \sin \theta \right).
\label{eq:QQbar_SG}
\eeq
When $\delta \epsilon = 0$, 
the Hamiltonian describes the $F=0$ (zero fermion) sector of 
the SUSY sine-Gordon QM.
Furthermore, for $\alpha = 0$, 
the ground state preserves SUSY and 
its wave function can be exactly determined as
\beq
\Psi^{(0)} = \exp \left( \frac{m}{2g^2} \cos \theta \right), \hs{10} 
H \Psi^{(0)} = 0.
\label{eq:Psi0_SG}
\eeq
For $\alpha \not = 0$, 
SUSY is spontaneously broken due to the topological term.

\subsection{Exact results around the SUSY and QES points}
Before discussing nonperturbative contributions 
in the path integral formalism, 
we derive some exact results 
for the expansion coefficients of the ground state energy 
around the SUSY and QES points.

\subsubsection{Expansion around the SUSY point}

\paragraph{Small $\delta \epsilon$ expansion \\}
First, we consider the expansion around the SUSY point ($\epsilon = 1, \alpha = 0$). 
The generating function is given by the modified Bessel function of the first kind
\beq
\langle 0 | 0 \rangle = \int_0^{2\pi} d \theta \, \exp \left( z \cos \theta \right) = 2\pi I_0(z), \hs{10}
z \equiv \frac{m}{g^2}. 
\eeq
The leading order Rayleigh-Schr\"odinger perturbation theory 
gives the first expansion coefficient of the ground state energy as
\beq
E^{(1)} = ~ \frac{\p}{\p \epsilon} E \bigg|_{\epsilon = 1} 
= \, - \frac{m}{2} \frac{\langle 0 | \cos \theta | 0 \rangle}{\langle 0 | 0 \rangle} 
\, = \, - \frac{m}{2} \frac{\p}{\p z} \log I_0(z). 
\label{eq:E1_SG}
\eeq
This is the exact result which will be compared with 
the semiclassical multi-bion contributions discussed below. 
To extract the perturbative contribution, let us consider
the weak coupling (large $z$) asymptotic expansion of $I_0(z)$: 
\beq
I_0(z) \ = \ \frac{e^{z}}{\sqrt{2\pi z}} 
\left[ 1 + \cdots  + \frac{1}{\pi} \frac{\Gamma(n+\frac{1}{2})^2}{\Gamma(n+1) \,} 
\left( \frac{1}{2z} \right)^n + \cdots \right] + \mathcal O(e^{-z}).
\eeq
This is the divergent power series 
which gives the asymptotic expansion of 
the generating function $\langle 0 | 0 \rangle$.  
Now let us consider the Borel resummation and 
see how the imaginary ambiguity arises from the perturbation series. 
First, replacing $\Gamma(n+\frac{1}{2})$ 
with its integral representation 
$\Gamma \left( n+\frac{1}{2} \right) = \int_0^\infty \frac{dt}{t} e^{-t} \, t^{n+\frac{1}{2}}$, 
we can rewrite the series as 
\begin{eqnarray}
\frac{e^{z}}{\sqrt{2\pi z}} \sum_{n=0}^\infty \frac{1}{\pi} 
\frac{\Gamma(n+\frac{1}{2})^2}{\Gamma(n+1) \,} \left( \frac{1}{2z} \right)^n 
&\to& 
\frac{e^{z}}{\sqrt{2\pi z}} 
\int_0^\infty dt ds \, e^{-t-s}\sum_{n=0}^\infty 
\frac{\ (ts)^{n+\frac{1}{2}}}{\pi\Gamma(n+1)} \left( \frac{1}{2z} \right)^n. 
\label{eq:test}
\end{eqnarray}
Then summing over $n$ before the integrations, 
we obtain the Borel resummation of the divergent series in $I_0$: 
\beq
\eqref{eq:test}
\, = \, \frac{1}{\pi} \int \frac{dt ds}{\sqrt{2\pi z ts}} \, e^{z-t-s+\frac{st}{2z}} 
\, = \, \frac{1}{\pi} \int_0^\infty dt \frac{e^{z-t}}{\sqrt{t(2z-t)}}. 
\label{eq:divseries1_SG}
\end{eqnarray} 
\begin{figure}[h]
\hs{-20}
\begin{minipage}{0.45\hsize}
\centering
\includegraphics[width=95mm]{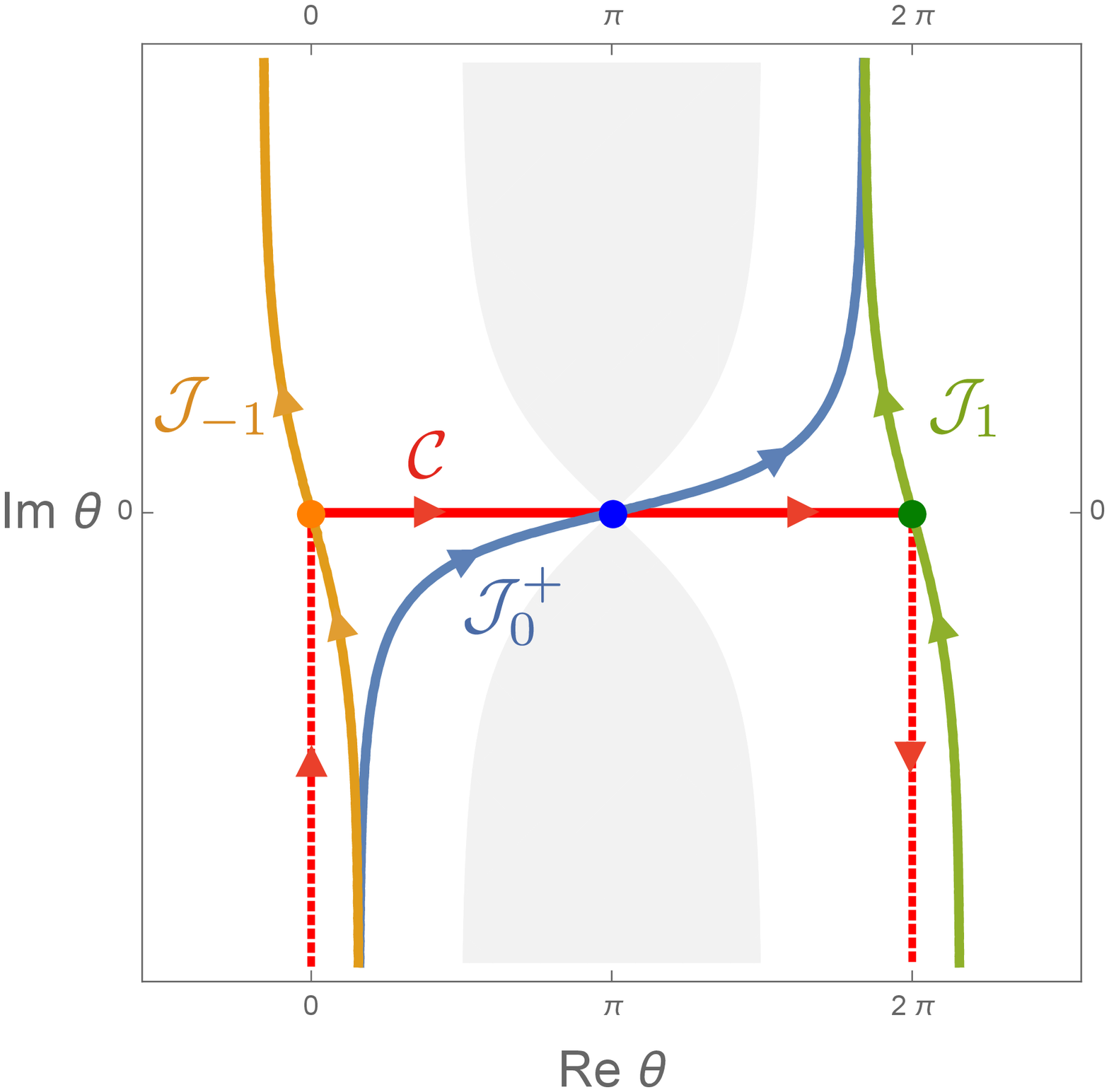} \\
\hs{15} (a) ~$\arg \, g^2 > 0$
\end{minipage}
\hs{5}
\begin{minipage}{0.45\hsize}
\includegraphics[width=95mm]{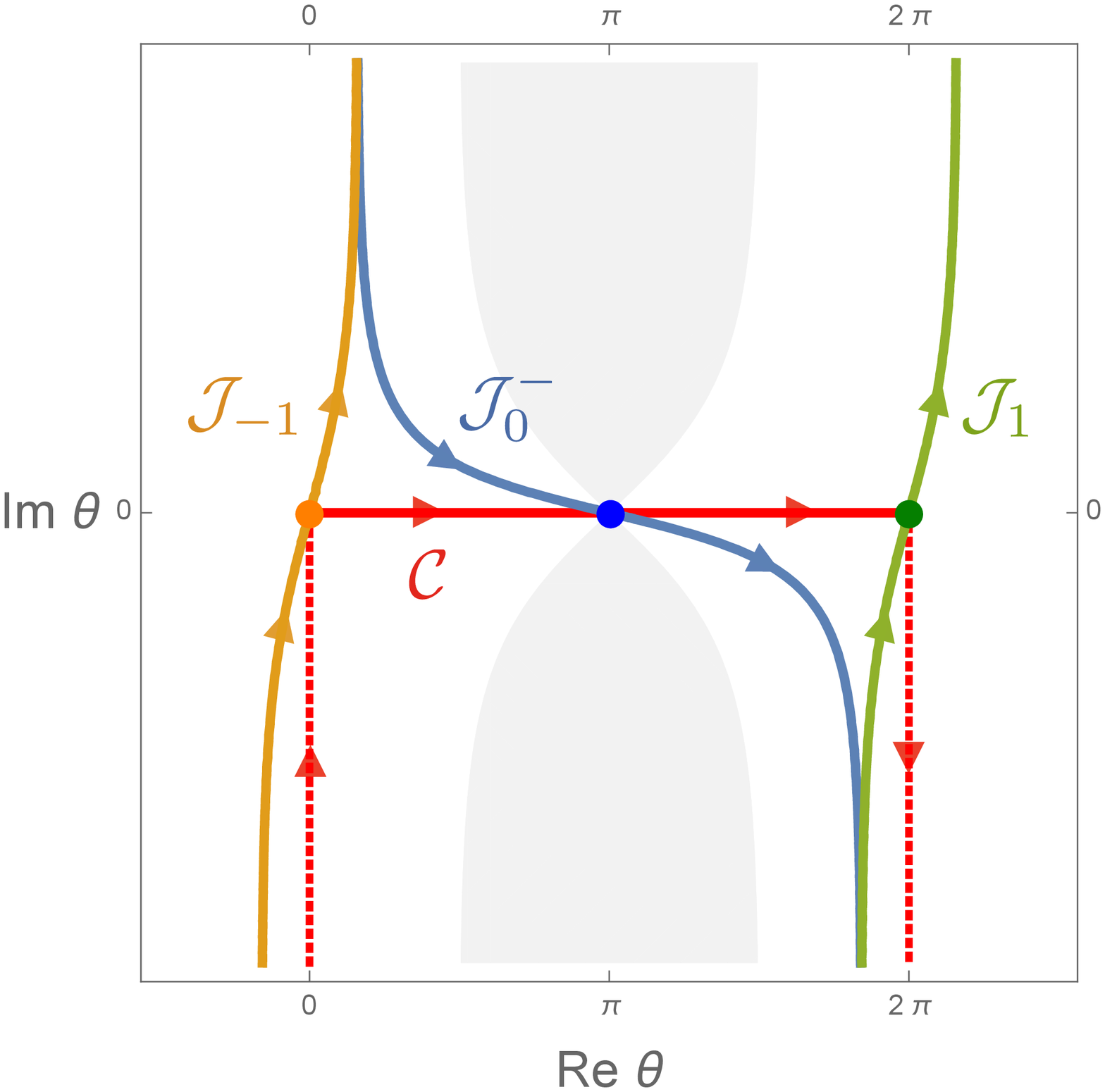} \vs{-6}  \\
\hs{15} (b) ~$\arg \, g^2 < 0$
\end{minipage}
\caption{The integration contour $C$ 
for the generating function $\langle 0 | 0 \rangle$ and
the Lefschetz thimbles $\mathcal J_n$ 
associated with the saddle points $\theta = (n+1)\pi$. 
The thimble with $n=0$ jumps at $\arg g^2 = 0$ 
due to the Stokes phenomenon.
The original integration contour $\mathcal C$ can be deformed 
and decomposed as $\mathcal J_0^+ - \mathcal J_1$ or $\mathcal J_0^- + \mathcal J_{-1}$ depending on $\arg g^2$. 
The ambiguous Borel resummation $I_0(z) \pm \frac{i}{\pi} K_0(z)$ corresponds to the integration along $\mathcal J_0^\pm$. }
\label{fig:Bessel_Stokes}
\end{figure}
Corresponding to the Stokes phenomenon at $\arg \, g^2=0$ 
shown in Fig.\,\ref{fig:Bessel_Stokes}, 
this integral representation has an imaginary ambiguity 
associated with the branch cut starting from $t=2z$ to infinity:
\beq
\frac{1}{\pi} \int_0^\infty dt \frac{e^{z-t}}{\sqrt{t(2z-t)}} \, 
= \, I_0(z) \pm \frac{i}{\pi} K_0(z),
\eeq
where $+$ is for $\arg z < 0$ and $-$ is for $\arg z >0$. 
Thus, we obtain the ambiguous perturbative part of $E^{(1)}$ as
\beq
E^{(1)}_0 \, = - \frac{m}{2} \frac{\p}{\p z} \log 
\left[\frac{1}{\pi} \int_0^\infty dt 
\frac{e^{z-t}}{\sqrt{t \left( 2z - t \right)}}\right] 
\ = - \frac{m}{2} \frac{\p}{\p z} \log \left[  I_0 (z) 
\pm \frac{i}{\pi} K_0(z) \right].
\label{E10_SUSY_SG}
\eeq
The remaining nonperturbative part, 
which cancel the imaginary ambiguity of the perturbative part, 
can be expressed as the convergent power series 
in the nonperturbative exponential $e^{-2z}$ as 
\beq
E^{(1)}-E^{(1)}_0  
\, = - \frac{m}{2} \frac{\p}{\p z} \log \left[1-\frac{\pm\frac{i}
{\pi} K_0(z)}{I_0 (z) \pm \frac{i}{\pi} K_0(z)} \right] 
\, = \, \sum_{p=1}^\infty E^{(1)}_p, 
\label{E10_SUSY_SG}
\eeq
where $E^{(1)}_p$ denotes the contribution 
with $p$-th power of nonperturbative exponential $e^{-2z}$
\beq
E^{(1)}_p = \ \frac{m}{2} \frac{\p}{\p z} 
\left[ \frac{1}{p} \left( \frac{\pm \frac{i}{\pi} K_0(z)}{I_0(z) \pm \frac{i}{\pi} K_0(z)} \right)^p \right] \, \sim \, \mathcal O(e^{-2pz}). 
\label{eq:p-th_order}
\eeq
This can be further expanded in powers of $g^2$ and 
the leading order term is given by
\beq
E^{(1)}_p = - m \left( \pm i e^{-\frac{2m}{g^2}} \right)^p + \cdots,
\label{eq:j=0}
\eeq
where we have used $K_0(z)/I_0(z) \approx \pi e^{-2z}$ for large $z$. 
This nonperturbative contribution with the imaginary ambiguity is expected to correspond to 
the semiclassical part of the $p$-bion contribution.
We will compare this exact result 
with the multi-bion contributions later.

\paragraph{Small $\alpha$ expansion \\}
It is also possible to obtain exact results 
by expanding the ground state energy in powers of $\alpha$ 
around the SUSY point $\epsilon=1$, $\alpha=0$.
The leading order correction $\Psi^{(1)}$ to the wave function 
can be determined from the expanded Schr\"odinger equation 
$\bar Q_0 Q_0 \Psi^{(1)} = - 2 i g^2 \p_\theta \Psi^{(0)}$ 
with $Q_0=Q \ (\alpha=0)$ and $\bar Q_0=\bar Q \ (\alpha=0)$,
which can be solved as 
\beq
\Psi^{(1)} = i \Psi^{(0)} \int_0^\theta d \theta' 
\left( 1 - \frac{e^{-z \cos\theta'}}{I_0(z)} \right). 
\label{eq:delpal_SUSY_SG}
\eeq
Since the Hamiltonian takes the form 
$H = \bar Q_0 Q_0 + 2 i \alpha g^2 \p_\theta + \alpha^2 g^2$, 
the ground state energy can be expanded as
\beq
E = \alpha \frac{\langle 0 | 2 i g^2 \p_\theta | 0 \rangle}
{\langle 0 | 0 \rangle} + \alpha^2 \left[ \frac{\langle 0 | 
g^2 | 0 \rangle}{\langle 0 | 0 \rangle} 
- \frac{\langle \Psi^{(1)} | \bar Q_0 Q_0 | \Psi^{(1)} \rangle}
{\langle 0 | 0 \rangle} \right] + \cdots. 
\eeq
Therefore, the first and the second order expansion 
coefficients are given by
\beq
\frac{\p}{\p \alpha} E \bigg|_{\alpha = 0} = 0, \hs{15}
\frac{1}{2} \frac{\p^2}{\p \alpha^2} E \bigg|_{\alpha = 0} = g^2I_0(z)^{-2}.
\label{eq: Ealal2_SUSY_SG}
\eeq
The second order expansion coefficient can be decomposed into the multi-bion contributions, 
whose $p$-bion part is given by
\beq
\frac{1}{2} \frac{\p^2}{\p \alpha^2} E_p \bigg|_{\alpha = 0} &=& 
g^2 p \frac{\left[ \pm \frac{i}{\pi} K_0(z) \right]^{p-1}}{\left[ I_0(z) \pm \frac{i}{\pi} K_0(z) \right]^{p+1}} 
\ = \mp 2 \pi i p m \left( \pm i e^{-\frac{2m}{g^2}} \right)^p + \cdots.
\label{eq:small_alpha}
\eeq
This shows that the nontrivial structure appears from the second order in the small $\alpha$ expansion.
We will later discuss these results in comparison with the multi-bion contributions.

\subsubsection{Expansion around the QES points}
In addition to the SUSY case, 
we can also obtain exact results in the QES case. 
To investigate the QES case, we rewrite the Hamiltonian as
\beq
H \ = \ \Psi^{(0)} \bigg[ g^2 J_3^2 - \frac{m}{2} (J_+ + J_-) \bigg] (\Psi^{(0)})^{-1},
\label{eq:H_QES_SG}
\eeq
where $J_3$ and $J_\pm$ are differential operators defined by
\beq
J_3 = i(\p_\theta - i \alpha), \hs{10} J_{\pm} = \mp e^{\mp i \theta} \big[ \p_\theta - i (\alpha \mp j) \big],
\eeq 
with $j \equiv \frac{\epsilon - 1}{2}$.
These operators satisfy the $\mathfrak{sl}(2)$ algebra
\beq
[J_+,J_-] = 2 J_3, \hs{5} [J_3,J_\pm]=\pm J_\pm, \hs{5} 
\mbox{with} \hs{5} J_3^2 + \frac{1}{2}(J_+ J_- + J_- J_+) = j(j+1). 
\label{eq:su2_SG}
\eeq
This $\mathfrak{sl}(2)$ algebra have 
a finite dimensional irreducible representation 
when $j$ is a nonnegative half integer. 
Note that eigenfunctions in such an irreducible representation satisfy 
the periodic boundary condition $\Psi(\theta + 2\pi) = \Psi(\theta)$ 
only when $j \pm \alpha$ is an integer 
(\textit{e.g.} the highest weight state $\Psi(\theta) = e^{i(\alpha-j)\theta} \Psi^{(0)}$).
Since the action of the Hamiltonian is closed 
within the irreducible representation, 
we can find its eigenfunctions using the ansatz
\beq
\Psi = \Psi^{(0)} \, (a_0 + a_1 J_- + \cdots + a_{2j} J_-^{2j} ) 
\, e^{i(\alpha-j)\theta}.
\label{eq:Psi1_QES_SG}
\eeq
Thus, by diagonalizing the Hamiltonian 
in this finite dimensional sector of the Hilbert space, 
we can obtain $2j+1$ exact eigenstates 
with the Bloch angle $2\pi\alpha=0$ (periodic $\Psi$) for $j\in {\mathbb Z}$ 
and $2\pi\alpha=\pi$ (anti-periodic $\Psi$) for $j \in {\mathbb Z}+1/2$.
For the singlet case $j = \alpha = 0$ ($\epsilon = 1$),
the eigenfunction of the Hamiltonian \eqref{eq:H_QES_SG} is given by
\beq
\Psi = \Psi_0, \hs{10} E = 0,
\label{eq:SUSY_QES_SG}
\eeq
which is consistent with the results 
we have already shown in the SUSY case.

For $j=\frac{1}{2}$ ($\epsilon=2$), $\alpha = \frac{1}{2}$, 
we find two eigenstates corresponding to the doublet representation
\beq
\Psi_\pm = \Psi_0 \, (1 \mp e^{i \theta}), \hs{10} 
E_\pm = \frac{1}{4} (g^2 \pm 2m).
\label{eq:PsiE_QES_SG}
\eeq
This indicates that 
the energy eigenvalues of the ground state $E_-$ 
and the first excited state $E_+$
do not receive any nonperturbative correction\cite{Kozcaz:2016wvy}. 
The small $\delta \epsilon = \epsilon - 1$ expansion gives 
the first order expansion coefficient of the ground state energy as 
\beq
E^{(1)} = \, \frac{\p}{\p \epsilon} E_- \bigg|_{j=\alpha=\frac{1}{2}} 
= \, - \frac{m}{2} \frac{\p}{\p z} \log \big[ I_0 (z) + I_1 (z) \big]. 
\label{eq:E1_QES_SG}
\eeq
As in the SUSY case, we can extract the perturbative part 
using the Borel resummation as 
\beq
E^{(1)}_{0} = - \frac{m}{2} \frac{\p}{\p z} \log \left[ I_0 (z) + I_1 (z) \pm \frac{i}{\pi} (K_0(z)-K_1(z)) \right]. 
\label{eq:E10_QES_SG}
\eeq
The remaining nonperturbative part can be decomposed 
into the $p$-bion contirbution, 
whose leading order part in the small $g^2$ expansion is given by 
\beq
E^{(1)}_p = \frac{m}{2} \frac{\p}{\p z} \left[ \frac{1}{p} 
\left( \frac{\pm \frac{i}{\pi} \left[ K_0(z) - K_1(z) \right]}
{I_0(z) + I_1(z) \pm \frac{i}{\pi} \left[ K_0(z) - K_1(z) \right]} 
\right)^p \right] 
= - m \left( \mp \frac{ig^2}{4m} e^{-\frac{2m}{g^2}} \right)^p 
+ \cdots,
\label{eq:j=1/2}
\eeq
where we have used $[K_0(z)-K_1(z)]/[I_0(z)+I_1(z)] \approx 
- \frac{\pi}{4z} e^{-2z}$ for large $z$. 
We will discuss this result in comparison 
with the multi-bion contributions later.


\subsection{Multi-bion solutions and semiclassical contributions}
In this subsection, 
we discuss the multi-bion contributions to the ground state energy
and show that they have imaginary ambiguities 
which are necessary for consistent resurgence structure 
of the sine-Gordon QM. 
We derive the semiclassical contributions 
from multi-bion saddle points 
by applying the Lefschetz thimble method. 
We will see that their expansion coefficients 
around the SUSY and QES points 
consistently reproduce those obtained 
from the exact results discussed in the previous subsection. 

\subsubsection{Multi-bion solutions}
\begin{figure}[t]
\centering
\includegraphics[width=120mm, viewport = 3 153 803 439]{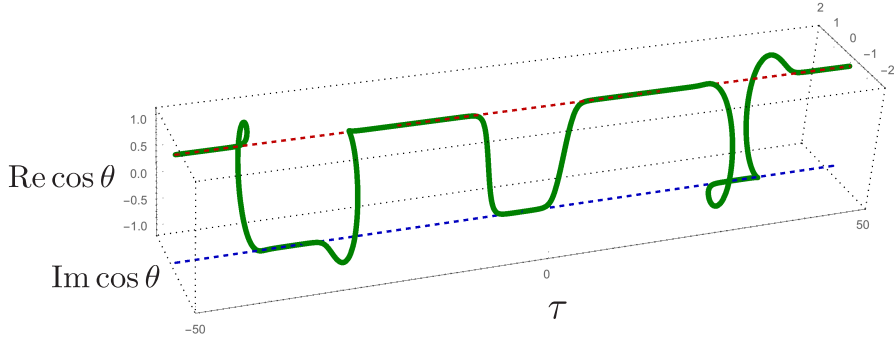} \hs{15}
\caption{Multi bion solution: $p=3$, $q=2$, $m=1$, $\epsilon=1$, $g^2=1/20000$, $\beta = 100$, $\tau_c=0$.}
\label{fig:multibion}
\end{figure}

We first identify the multi-bion solutions 
corresponding to the complex saddle points of 
the Euclidean action of the sine-Gordon QM \eqref{eq:Salpha_SG}. 
They can be easily obtained from those in the $\C P^{1}$ QM \cite{Fujimori:2017oab} by ignoring the azimuthal angle variable.
The solutions of the Euclidean equation of motion. 
which have nontrivial contributions 
in the $\beta \to \infty$ limit take the form
\beq
\tan {\theta(\tau)\over{2}} = \frac{f(\tau-\tau_c)}{\sin \alpha}, 
\label{eq:sol_SG}
\eeq 
where $\tau_c$ is the complexified position modulus and $f(\tau)$ 
is the elliptic function 
\beq
f(\tau) = {\rm cs} (\Omega \tau, k) \equiv  
{\rm cn} (\Omega \tau, k)/ {\rm sn} (\Omega \tau, k), 
\label{eq:ftau_SG}
\eeq
which satisfies the differential equation 
$(\p_\tau f)^2 = \Omega^2 (f^2+1)(f^2+1-k^2)$. 
Since the periods of the doubly periodic function ${\rm cs}(z,k)$ 
are given by the elliptic integrals $2K(k)$ and 
$4i K'(k) \equiv 4i K(\sqrt{1-k^2})$, 
the periodic boundary condition 
$\theta(\tau+\beta) = \theta(\tau)$ is satisfied 
when the parameters $(\Omega,k)$ are related to 
two integers $(p,q)$ as
\beq
\beta = \frac{2pK+4iqK'}{\Omega}, \hs{10} 0 \leq q < p. 
\eeq  
The parameters $(\alpha, \Omega, k)$ are 
given in terms of the period $\beta$ and the pair of integers $(p,q)$, 
and their asymptotic forms for large $\beta$ are given by 
(see Appendix B of \cite{Fujimori:2017oab} for details)
\beq
&k \approx 1 - 8 \, e^{ - \frac{\omega \beta - 2 \pi i q}{p}}, \hs{3}
\Omega \approx \omega \left( 1 + 8 \frac{\omega^2+m^2}{\omega^2-m^2} 
e^{ - \frac{\omega \beta - 2 \pi i q}{p}} \right), & \notag \\
&\cos \alpha \approx \frac{m}{\omega} \left( 1 - \frac{8m^2}{\omega^2-m^2} 
e^{ - \frac{\omega \beta - 2 \pi i q}{p}} \right) , &
\label{eq:parameters_SG}
\eeq
where $\omega = m \sqrt{1+ \epsilon g^2/m}$.  
The asymptotic value of the action 
for the $(p,q)$ solution takes the form
\beq
S \, \approx \, p S_1 + \pi i \epsilon l ~~~\mbox{for large $\beta$}, 
\label{eq:Baction_SG}
\eeq
where $S_1$ denotes the on-shell value of the action 
for the single bion configuration
\beq
S_1 = \frac{2m}{g^2} + \epsilon \log \frac{\omega + m}{\omega - m},
\eeq
and we have ignored the vacuum value of the acition. 
The imaginary part $\pi i \epsilon l$ is related to 
the so-called hidden topological angle \cite{Behtash:2015kna} 
and the integer $l$ is zero or 
the greatest common divisor of $p$ and $2q$ 
depending on the value of ${\rm Im} \, \tau_c$.
The real part of the action is 
$p$ times the single bion action $S_1$,
which shows that the integer $p$ is the number of bions. 
We can see from the solution \eqref{eq:sol_SG} that 
the bions are equally spaced 
and $n$-th kink and antikink are located 
at $\tau_n^+$ and $\tau_n^-$ given by 
\beq
\tau_n^\pm \, = \, \tau_c + \frac{n-1}{\omega p} (\omega \beta - 2\pi i q) 
\pm \frac{1}{2\omega}\log {\frac{4\omega^2}{\omega^2-m^2}}.
\label{eq:QM_saddle}
\eeq
Fig.\,\ref{fig:multibion} shows an example of 
a multi-bion solution with $(p,q)=(3,2)$. 

\subsubsection{Multi-bion contributions and Quasi-moduli integral}
Next, we compute the semiclassical contributions 
from the multi-bion saddle points 
in the weak coupling limit $g \rightarrow 0$.  
As can be seen from Eq.\,\eqref{eq:QM_saddle}, 
each constituent (anti)kink is almost isolated 
for large $\beta$ and small $g$
and hence the inter-kink binding force becomes small. 
Therefore, the positions of the constituent (anti)kinks can be 
regarded as quasi-moduli parameters, 
which parametrize the nearly-flat directions 
around the multi-bion saddle points. 
As explained in Appendix \ref{appendix:quasi_moduli}, 
the semiclassical contribution from the saddle points
can be obtained by reducing the path integral  
to a finite dimensional integral over the quasi-moduli parameters.

Let us consider the semiclassical contributions 
from the $p$-bion saddle points. 
There are $2p$ constituent kinks
and each of them can be either instanton or anti-instanton. 
Assigning $s_i= +1~(-1)$ if $i$-th kink is an (anti-)instanton,
we can write the $p$-bion contribution 
to the partition function $Z_p$ as
\beq
\frac{Z_p}{Z_0} = \sum_{s_1=\pm 1} \cdots \sum_{s_{2p} = \pm 1} X_{(s_1,\cdots,s_{2p})},
\label{eq:ZpZ0_SG}
\eeq
where $Z_0$ is the perturbative part of the partition function and 
$X_{(s_1, \cdots, s_{2p})}$ denotes the quasi-moduli integral 
for a fixed set of $s_i$: 
\beq 
X_{(s_1, \cdots, s_{2p})} = \frac{1}{p} e^{-\frac{2pm}{g^2}} 
\int \prod_{i=1}^{2p} \left[ m d\tau_i \sqrt{\frac{2m}{\pi g^2}} \exp \left( - V_i \right) \right]. 
\label{eq:X_SG}
\eeq
The factor $\sqrt{\frac{2m}{\pi g^2}}$ is 
the one-loop determinant around each kink
and $1/p$ is inserted since the kinks are indistinguishable. 
The nearest-neighbor interaction potential $V_i$ 
between the $i$-th and $(i+1)$-th kinks, 
which is discussed in Appendix \ref{appendix:effpot}, 
is given by 
\beq
V_i = \frac{4m}{g^2} s_{i} s_{i-1} e^{-y_i} + \epsilon_i y_i - \pi i \alpha s_i, \hs{10} 
y_i = m(\tau_{i}-\tau_{i-1}) .
\label{eq:Vi_SG}
\eeq
where $\tau_0 = \tau_{2p}$, $s_0 = s_{2p}$ and 
\beq
\epsilon_i = \left\{ \begin{array}{cc} \epsilon & \mbox{for even $i$} \\ 0 & \mbox{for odd $i$} \end{array} \right..
\label{eq:epi_SG}
\eeq
We can check that the saddle points of 
the effective potential $\sum_{i=1}^{2p} V_i$
give complexified kink positions which are consistent
with those read off from the multi-bion solution \eqref{eq:QM_saddle}. 

To evaluate the quasi-moduli integral, 
it is convenient to introduce the Lagrange multiplier for the constraint
\beq
\delta \left( \sum_{i=1}^{2p} \tau_i - \beta \right) = 
m \int_{-\infty}^{\infty} \frac{d \sigma}{2\pi} \exp \left[ i m \sigma \left( \sum_{i=1}^{2p} \tau_i - \beta \right) \right]. 
\label{eq:LagrangeM_SG}
\eeq
Then, we can rewrite $X_{(s_1, \cdots, s_{2p})}$ as 
\beq
X_{(s_1, \cdots, s_{2p})} = \frac{m \beta}{p} e^{-\frac{2pm}{g^2}} \int \frac{d\sigma}{2\pi} e^{-i m \beta \sigma} \prod_{i=1}^{2p} I_i,
\label{eq:X2_SG}
\eeq
where $I_i$ is given by the following integral with respect to the single valuable $\tau_i$: 
\beq
I_i \ = \ \sqrt{\frac{2m}{\pi g^2}} \int m d\tau_i \, 
\exp \left( -\frac{4m}{g^2} s_{i} s_{i-1} e^{-m \tau_i} - 
(m \epsilon_i - i \sigma) \tau_i + \pi i \alpha s_i \right),
\label{eq:Ii_SG}
\eeq
where the integration contour is determined 
by the Lefschetz thimble method. 
Since the details of the thimble calculation are parallel to 
those for the quasi moduli integral 
in the $\C P^{1}$ QM in \cite{Fujimori:2017oab}, 
we show only the essential parts of the calculation below. 

As shown in Appendix \ref{appendix:QMI}, 
we can evaluate $I_i$ by means of the Lefschetz thimble method as
\beq
I_i \ = \ \sqrt{\frac{2m}{\pi g^2}} \left( \frac{4m}{g^2} \right)^{i \sigma - \epsilon_i}  \Gamma \left( \epsilon_i - i \sigma \right) 
\exp \left[ \pm \frac{\pi i}{2} ({i \sigma - \epsilon_i}) (1- s_i s_{i-1}) + \pi i \alpha s_i \right],
\label{eq:Ii2_SG}
\eeq
where the ambiguous sign comes from the Stokes phenomenon: 
the sign $\pm$ corresponds to the limit $\arg\,g^{2} \rightarrow \pm 0$. 
\begin{figure}[t]
\includegraphics[width=100mm]{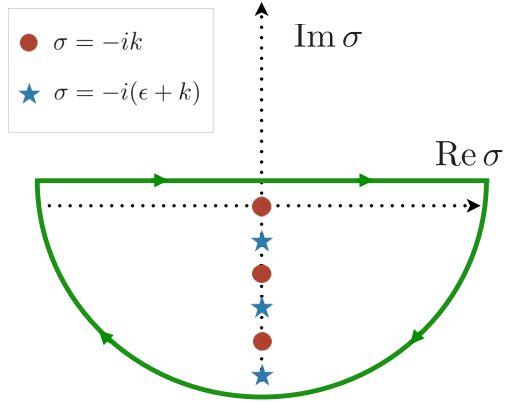}
\caption{Integration contour for $\sigma$. 
The poles of the integrand are located at $\sigma = -i k$ and $\sigma=-i(\epsilon+k)~(k \in \Z_{\geq 0})$.}
\label{fig:sigma_contour}
\end{figure}
To obtain this expression, 
we have shifted ${\rm Im} \, \sigma$ so that ${\rm Re}(\epsilon_i - i \sigma) > 0$ for all $i$.
By closing the integration contour for $\sigma$ as shown in Fig.\,\ref{fig:sigma_contour}, 
$X_{(s_1, \cdots, s_{2p})}$ can be evaluated by picking up the poles of the integrand located at 
\beq
\sigma = - i k ~~~\mbox{and}~~ -i(\epsilon+k), \hs{10} k \in \Z_{\geq 0}.
\eeq 
Since the residues at $\sigma \not = 0$ vanishes in the $\beta \rightarrow \infty$ limit, 
the pole at $\sigma = 0$ gives the leading order term for large $\beta$ 
\beq
X_{(s_1, \cdots, s_{2p})} \ = - \frac{i m\beta}{p} e^{-\frac{2pm}{g^2}} \, \underset{\sigma = 0}{\rm Res} \left[ e^{-i m \beta \sigma} \prod_{i=1}^{2p} I_i \right] + \mathcal O(e^{-m \epsilon \beta}, e^{-m \beta}).
\label{eq:X3_SG}
\eeq
Thus we obtain the following semiclassical contribution of the $p$-bion solutions: 
\beq
\frac{Z_p}{Z_0} \approx 
- \frac{i m\beta}{p} e^{-\frac{2pm}{g^2}} \underset{\sigma = 0}{\rm Res} 
\Bigg[ e^{-i m\beta \sigma} \bigg\{ \frac{1}{2\pi} \left( \frac{4m}{g^2} \right)^{2i\sigma - \epsilon+1} \Gamma(-i\sigma) \Gamma(\epsilon-i\sigma) \bigg\}^p \mathcal Z_\pm \Bigg],
\label{eq:ZpZ02_SG}
\eeq
where $\mathcal Z_\pm$ is given by
\beq
\mathcal Z_\pm = \sum_{s_1=\pm 1} \cdots \sum_{s_{2p} = \pm 1} \prod_{i=1}^{2p} 
\exp \left[ \pm \frac{\pi i}{2} ({i \sigma - \epsilon_i}) (1- s_i s_{i-1}) + \pi i \alpha s_i \right] .
\label{eq:mathZ_SG}
\eeq
By using the transfer matrix 
\beq
T_\pm = \ba{cc} e^{\pi i \alpha}  & e^{\pm \pi i(i \sigma - \epsilon)} \\ e^{\pm \pi i(i \sigma - \epsilon)} & e^{-\pi i \alpha}  \ea 
\ba{cc} e^{\pi i \alpha}  & e^{\pm \pi i (i \sigma)} \\ e^{\pm \pi i(i \sigma)} & e^{-\pi i \alpha}  \ea, 
\label{eq:Tmat_SG}
\eeq
$\mathcal Z_\pm$ can also be written as 
\beq
\mathcal Z_\pm ~=~ \tr (T_\pm^p).
\label{mathZ2_SG}
\eeq
By calculating $\tr (T_\pm^p)$ and evaluating the residue of Eq.\,\eqref{eq:ZpZ02_SG} at $\sigma = 0$, 
we can obtain the explicit $p$-bion contribution to the partition function. 

\subsubsection{Comparison with the exact results}
We are now ready to compare 
the semiclassical contributions from the multi-bion solutions 
with the exact results (\ref{eq:j=0}), (\ref{eq:small_alpha}), and (\ref{eq:j=1/2}). 
We show below that the semiclassical multi-bion contributions obtained above 
precisely agree with the exact results 
for the expansion coefficients of the ground state energy around the SUSY and QES points. 
\vs{5} 
\paragraph{Small $\delta \epsilon$ expansion around $j \in \Z$, $\alpha = 0$ \\} 
The trace $\tr (T_\pm^p)$ can be expanded around 
the QES point $\epsilon = 2j+1~(j \in \Z)$, $\alpha = 0$ as
\beq
\tr (T_\pm^p) &=& 2^p e^{\mp p \pi \sigma} \Big[ 2 \sinh^p ( \pm \pi \sigma ) 
\pm i p \pi e^{\mp \pi \sigma} \sinh^{p-1} (\pm \pi \sigma) \, \delta \epsilon 
+ \mathcal O (\delta \epsilon^2) \Big],
\label{eq:Tp_al0_SG}
\eeq
where $\delta \epsilon = \epsilon-2j-1~(\forall j \in \Z)$. 
The first term indicates that 
the trace $\tr (T_\pm^p)$ has a $p$-th order zero 
at $\sigma = 0$ for $\delta \epsilon = 0$,
so that the $p$-bion contribution \eqref{eq:ZpZ02_SG} vanishes at the QES point
\beq
E^{(0)}_p = \lim_{\beta \rightarrow \infty} \lim_{\delta \epsilon \rightarrow 0} \left( - \frac{1}{\beta} \frac{Z_p}{Z_0} \right) = 0.
\label{eq:E0p_al0_SG}
\eeq 
The second term in Eq.\,\eqref{eq:Tp_al0_SG} gives 
the following contribution to the first expansion coefficient of the ground state energy:  
\beq
E^{(1)}_p = \lim_{\beta \rightarrow \infty} \lim_{\delta \epsilon \rightarrow 0} \frac{\p}{\p \delta \epsilon} \left( - \frac{1}{\beta} \frac{Z_p}{Z_0} \right) = - m \left[ \pm i \frac{\Gamma(2j+1)}{(4z)^{2j}} e^{-\frac{2m}{g^2}} \right]^p + \cdots.
\label{eq:E1p_al0_SG}
\eeq
This is consistent with the exact result for $j=0$ \eqref{eq:j=0}, 
which indicates that the imaginary ambiguity 
from the perturbation series is canceled 
by those from the semiclassical bion contributions. 

\vs{5}
\paragraph{Small $\alpha$ expansion around $j \in \Z$, $\alpha = 0$ \\}
Next, let us consider the expansion with respect to $\alpha$. 
For $\epsilon =2 j + 1$, the trace $\tr (T_\pm^p)$ can be expanded as 
\beq
\tr (T_\pm^p) &=& 2^p e^{\mp p \pi \sigma} 
\Big[ 2 \sinh^p ( \pm \pi \sigma ) + 2 p^2 \pi^2 e^{\mp \pi \sigma} \sinh^{p-1} (\pm \pi \sigma) \, \alpha^2 + \mathcal O (\alpha^3) \Big].
\label{eq:Tp_al_SG}
\eeq 
The absence of an $\mathcal O(\alpha)$ term implies that 
the leading order expansion coefficient vanishes 
\beq
\frac{\p E_p}{\p \alpha} \bigg|_{\alpha=0} = 0. 
\eeq
The first nontrivial contribution appears 
from the second order coefficient 
\beq
\frac{1}{2} \frac{\p^2 E_p}{\p \alpha^2} \bigg|_{\alpha=0} = \lim_{\beta \rightarrow \infty} \lim_{\alpha \rightarrow 0} \frac{1}{2} \frac{\p^2}{\p \alpha^2} \left( - \frac{1}{\beta} \frac{Z_p}{Z_0} \right) = \mp 2 \pi i p m \left[ \pm i \frac{\Gamma(2j+1)}{(4z)^{2j}} e^{-\frac{2m}{g^2}} \right]^p + \cdots.
\label{eq:Ealal_bion_SG}
\eeq
These results are consistent with 
those obtained from the exact expressions 
for $j=0$ given in Eqs.\,\eqref{eq: Ealal2_SUSY_SG} and \eqref{eq:small_alpha}. 

\vs{5}
\paragraph{Small $\delta \epsilon$ expansion around $j \in \Z+\frac{1}{2}$, $\alpha = \frac{1}{2}$ \\}
Finally, let us look into the expansion around $j \in \Z + \frac{1}{2}$, $\alpha = \frac{1}{2}$. 
The trace $\tr (T_\pm^p)$ can be expanded as 
\beq
\tr (T_\pm^p) &=& 2^p e^{\mp p \pi \sigma} 
\Big[ 2 \sinh^p ( \mp \pi \sigma ) 
\mp i p \pi e^{\mp \pi \sigma} \sinh^{p-1} (\mp \pi \sigma) \, \delta \epsilon + \mathcal O (\delta \epsilon) \Big],
\label{eq:Tp_j_SG}
\eeq
where $\delta \epsilon = \epsilon-2j-1~(\forall j \in \Z + \frac{1}{2})$.
As above, we can obtain 
the following semiclassical $p$-bion contribution 
to the expansion coefficients, 
which is consistent with the exact result 
\eqref{eq:j=1/2} for $j=\frac{1}{2}$: 
\beq
E^{(0)}_p = 0, \hs{10} E^{(1)}_p = - m \left[ \mp i \frac{\Gamma(2j+1)}{(4z)^{2j}} e^{-\frac{2m}{g^2}} \right]^p + \cdots.
\label{eq:E0p_j_SG}
\eeq

We have shown in the sine-Gordon QM that 
the multi-bion semiclassical contributions correctly reproduce 
the leading (semiclassical) terms of the exact results 
for the ground state energy
around the SUSY and QES points in the parameter space. 
We emphasize that this agreement reveals the resurgence structure 
to all orders of the nonperturbative exponential, 
even though the resurgence structure 
in models such as the sine-Gordon model 
is more complicated than that of the K\"ahler QM 
such as the ${\mathbb C}P^{N-1}$ QM
discussed in the next section.


\section{Resurgence Structure in $\C P^{N-1}$ quantum mechanics}
\label{sec:CPN}

In this section, we consider quantum mechanics 
on the complex projective space $\C P^{N-1}$ 
as an example of a model on a K\"ahler target space 
discussed in Sec.\,II. 

\subsection{$\C P^{N-1}$ quantum mechanics}

The K\"ahler potential $K$ and K\"ahler metric $G_{i\bar j}$ of 
$\C P^{N-1}$ are given by 
\beq
K = \log (1 + |\varphi^i|^2), \hs{10}
G_{i \bar j} = \frac{\p^2 K}{\p \varphi^i \p \bar \varphi^{\bar j}} 
= \frac{(1+|\varphi^i|^2) \delta_{i \bar j} 
- \varphi^{j} \bar \varphi^{\bar i}}{(1+|\varphi^i|^2)^2} \,.
\label{eq:CPN_KG}
\eeq
This K\"ahler metric has an $SU(N)$ holomorphic isometry 
and we use the following specific linear combination of the Killing vectors in its Cartan subalgebra: 
\beq
\Xi = -\sum_{i=1}^{N-1} i m_i ( \varphi^i \p_i - \bar \varphi^{\bar i} \bar \p_{\bar i} ), \hs{10} 
\mu = \sum_{j=1}^{N-1} \frac{m_j |\varphi^j|^2}{1+|\varphi^i|^2} \,.
\label{CPN_Ximu}
\eeq
The coefficients $m_i$, 
which parametrize the Killing vector $\Xi$ and 
the moment map $\mu$,
determine the potential in the Sch\"odinger equation of 
the $\C P^{N-1}$ QM
\beq
H \Psi = G^{\bar j i} \left[ - g^2 \p_i \bar \p_{\bar j} + \frac{1}{g^2} \p_i \mu \, \bar \p_{\bar j} \mu - \epsilon \p_i \bar \p_{\bar j} \mu \right] \Psi = E \Psi\,.
\label{eq:CPNSch1}
\eeq
For $\epsilon=1$, this equation describes the SUSY $\C P^{N-1}$ QM
projected to the sector with the lowest fermion number $F$. 
The SUSY ground state wave function and its energy eigenvalue are given by
\beq
\Psi^{(0)} = \langle \varphi | 0 \rangle = \exp \left( - \frac{\mu}{g^2} \right), \hs{10} 
E^{(0)} = 0.
\label{eq:Psi0E0CPN}
\eeq
For $\epsilon \approx 1$, 
we can solve the Schr\"odinger equation  
by expanding the wave function and the ground state energy 
with respect to $\delta \epsilon = \epsilon - 1 $ as
\beq
\Psi =\Psi^{(0)}+ \delta \epsilon \Psi^{(1)} + \cdots, \hs{10} 
E = \delta \epsilon \, E^{(1)} + \delta \epsilon^2 E^{(2)} + \cdots. 
\label{eq:ExpanCPN}
\eeq 
These expansion coefficients are determined by the standard Rayleigh-Schr\"odinger perturbation theory
\beq
E^{(1)} = - \frac{\langle 0 | \Delta \mu | 0 \rangle}{\langle 0 | 0 \rangle},
\hs{7} 
E^{(2)} = - \frac{\langle \Psi^{(1)}| H_{\epsilon=1} | \Psi^{(1)}\rangle}{\langle 0 | 0 \rangle}, 
\hs{7} \cdots,
\label{eq:coefficientsCPN}
\eeq
where we have used $\delta H = -\Delta\mu$. 
$\langle 0 | 0 \rangle$ is the normalization factor of 
the SUSY ground state wave function in Eq.\,\eqref{eq:Psi0E0CPN}, 
which can also be viewed as the generating function for $\mu$:
\beq
\langle 0 | 0 \rangle = \int dv \, \exp \left( - \frac{2\mu}{g^2} \right), \hs{10} 
dv = \frac{1}{(N-1)!} \left( \frac{i}{2} G_{i \bar j} d\varphi^i \wedge d \bar \varphi^{\bar j} \right)^{N-1}. 
\eeq 
For $N=2$ (the $\C P^{1}$ case), 
the resurgence structure to all orders with complex multi-bion solutions 
have been investigated in our previous work \cite{Fujimori:2017oab}. 
Here we discuss the case with general $N \geq 2$.


\subsection{Near-SUSY exact results}
First let us derive some exact results 
for the expansion coefficients $E^{(n)}$ 
around the SUSY point $\epsilon =1$. 
Since the perturbation Hamiltonian $\delta H = -\Delta\mu$ 
in this case is simply given by the following linear function of $\mu$ 
\beq
- \Delta\mu \, = \, G^{\bar j i} \bar \p_{\bar j} \p_i \mu 
\, = \, - \sum_{i=1}^{N-1} m_i + N \mu \,,
\label{eq:DmuCPN}
\eeq
the first expansion coefficient can be rewritten as
\beq
E^{(1)} \, = \, - \frac{\langle 0 | \Delta \mu | 0 \rangle}
{\langle 0 | 0 \rangle} 
\, = \, - \sum_{i=1}^{N-1} m_i + \frac{N}{2} g^4 \frac{\p}{\p g^2} 
\log \langle 0 | 0 \rangle.
\label{eq:00E1}
\eeq
As shown in Appendix \ref{app:Loc}, 
the generating function $\langle 0 | 0 \rangle$ can be calculated 
by the Duistermaat-Heckman localization formula \cite{Atiyah-Bott, DT} as 
\beq
\langle 0 | 0 \rangle ~=~ \left( \prod_{i=1}^{N-1} \frac{\pi g^2}{2m_i} \right) \left(1 - \sum_{i=1}^{N-1} A_i e^{-\frac{2m_i}{g^2}} \right), 
\label{eq:normfacCPN}
\eeq
where the coefficients $A_i$ are given by
\beq
A_i = {\prod_{j = 1,\, j \not=i}^{N-1}} \frac{m_j}{m_j-m_i}.
\label{eq:constant-Ai}
\eeq 
Combining Eqs.\,\eqref{eq:00E1}--\eqref{eq:constant-Ai}, we 
obtain the following exact expression
for the first order expansion coefficient $E^{(1)}$: 
\beq
E^{(1)} &=& \frac{N(N-1)}{2} g^2 - \sum_{i=1}^{N-1} m_i \left( 1 + \frac{N A_i e^{-\frac{2m_i}{g^2}}}{1 - \sum_{j=1}^{N-1} A_j e^{-\frac{2m_j}{g^2}}} \right).
\label{eq:CPN_E1}
\eeq
As mentioned in Sec.\,II, 
no divergent asymptotic series appears in each sector of the trans-series. 
For example, the perturbative contribution is given by
\beq
E^{(1)}_{0} = \frac{N(N-1)}{2} g^2 - \sum_{i=1}^{N-1} m_i \,.
\label{eq:E10_CPN}
\eeq
The absence of divergent asymptotic series indicates 
that the first order expansion coefficient $E^{(1)}$ has no nontrivial resurgence structure 
among the sectors with different orders of nonperturbative exponentials. 

Next, let us move on to the second order expansion coefficient of the ground state energy $E^{(2)}$.
To obtain $E^{(2)}$ from Eq.\,\eqref{eq:coefficientsCPN}, 
we need to solve the $\mathcal O(\delta \epsilon)$ Schr\"odinger equation for $\Psi^{(1)}$. 
Although it is difficult to obtain the exact solution for $\Psi^{(1)}$, 
the Bender-Wu method \cite{BenderWu} can be used 
to determine $\Psi^{(1)}$ in a perturbative way. 
Resumming the perturbative series of $\Psi^{(1)}$, 
we obtain the following leading order part of $\Psi^{(1)}$ in the weak coupling limit $g \rightarrow 0$: 
\beq
\Psi^{(1)} = \frac{N}{2} e^{-\frac{\mu}{g^2}} \log \frac{1}{1+\sum_{k=1}^{N-1}|\varphi^k|^2} + \cdots, 
\label{eq:Psi1_CPN}
\eeq
where $\cdots$ denotes nonperturbative corrections. 
We can also directly check that Eq.\,\eqref{eq:Psi1_CPN} is the correct perturbative part of $\Psi^{(1)}$
by substituting it into the $\mathcal O(\delta \epsilon)$ Schr\"odinger equation. 

\begin{figure}[t]
\includegraphics[width=100mm, viewport = 124 244 992 597]{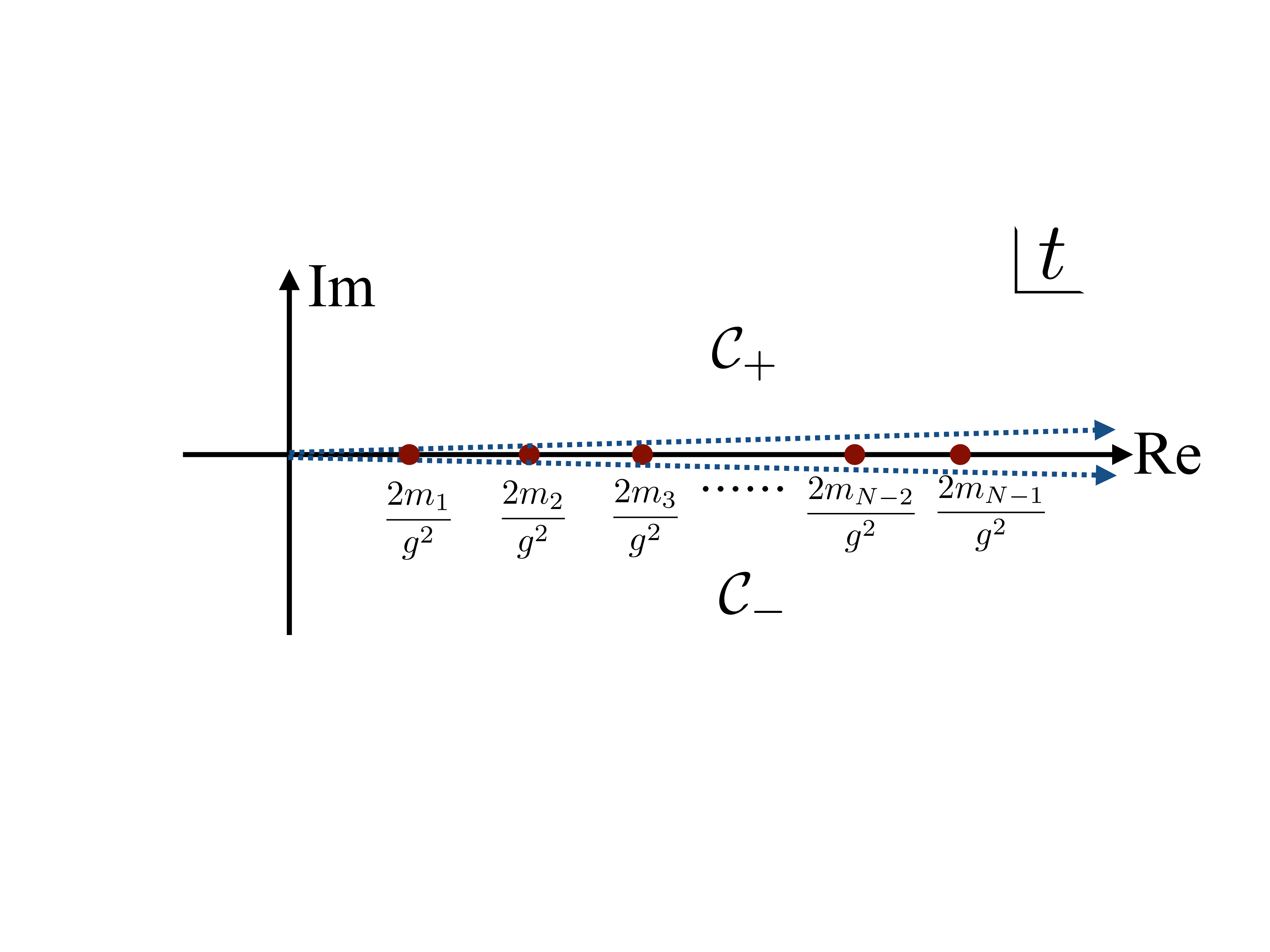}
\caption{Singularities on the Borel plane for the perturbative Borel transform of $\C P^{N-1}$ QM.}
\label{fig:multi-sing}
\end{figure}

Then, from the relation $E^{(2)} = - \langle \Psi^{(1)}| H_{\epsilon=1} | \Psi^{(1)}\rangle /\langle 0 | 0 \rangle$, 
we obtain the perturbative part of the second order expansion coefficient as (see Appendix \ref{appendix:E2CPN})
\beq
E^{(2)}_0 \, = \, \frac{N^2}{4} \left[ g^2 + \sum_{i=1}^{N-1} 2 m_i A_i \int_0^\infty dt \frac{e^{-t}}{t-\frac{2m_i}{g^2 \pm i 0}} \right].
\label{eq:CPN_E2}
\eeq
This perturbative part has the following imaginary ambiguity 
due to the regularization $g^2 \rightarrow g^2 \pm i 0$:
\beq
{\rm Im} \, E^{(2)}_0 = \mp  \frac{\pi i}{2} N^2 \sum_{i=1}^{N-1} m_i A_i e^{-\frac{2m_i}{g^2}}.
\label{eq:CPN_E2_amb}
\eeq
It is notable that the number of singularities on the Borel plane is $N-1$, 
which means that there are multiple singularities for $N>2$ as shown in Fig.\,\ref{fig:multi-sing}. 
We will show that the imaginary ambiguities 
which originate from these $N-1$ singularities are 
canceled by the $(N-1)$-type single (real and complex) bions.

\subsection{Bion solutions and semiclassical contributions} 
In the previous section, we have derived the imaginary ambiguities 
arising from the perturbation series of $E^{(2)}$. 
In this section, we show that they are canceled 
by the semiclassical contributions from single bion solutions. 
We also show that the contributions from multi-bion solutions 
correctly reproduce $E^{(1)}$ 
to all orders in the nonperturbative exponentials.

\subsubsection{Embedding $\C P^1$ bion solutions}
\begin{figure}[t]
\includegraphics[width=170mm, viewport = 39 201 859 392]{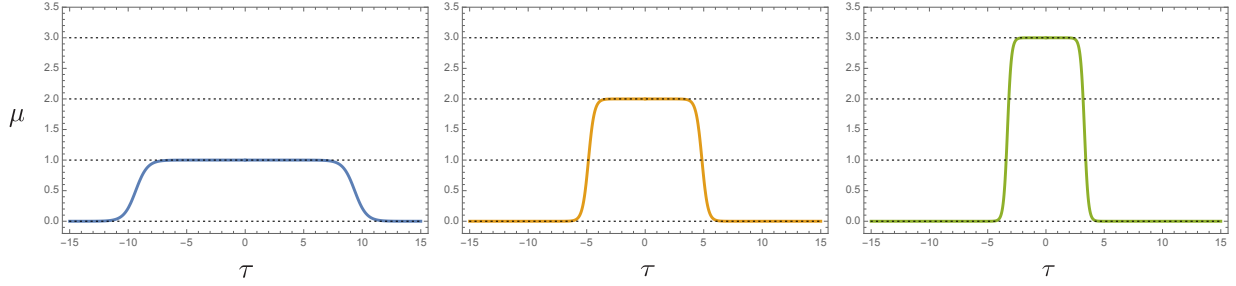}
\caption{Examples of $(N-1)$ bion solutions for $N=4$: $m_1=1$, $m_2=2$, $m_3=3$, $g=10^{-4}$, $\epsilon=1$.}
\label{fig:single-bions}
\end{figure}

First, we investigate exact (complex) saddle point solutions 
in the $\C P^{N-1}$ QM.
Here we focus on the $\beta \rightarrow \infty$ limit for simplicity.  
The Euclidean action for the projected $\C P^{N-1}$ QM takes the form 
\beq
S_E = \frac{1}{g^2} \int_{-\infty}^{\infty} d\tau \left[ G_{i \bar j} 
\p_\tau \varphi^i \p_\tau \bar \varphi^{\bar j} + 
G^{\bar j i} \left( \p_i \mu \, \bar \p_{\bar j} \mu - g^2 
\epsilon \, \p_i \bar \p_{\bar j} \mu \right) \right]. 
\label{eq:EucS_CPN}
\eeq
To find the simplest saddle point solution, 
let us consider the ansatz 
\beq
\varphi^j = 0 ~~\mbox{for}~~ j \not = 1. 
\eeq
Then the equation of motion reduces to that 
in the $\C P^1$ QM \cite{Fujimori:2017oab} with 
$2 \epsilon \rightarrow N \epsilon$,
and hence we can embed the $\C P^1$ bion solutions into $\varphi^1$ as 
\beq
\varphi^1 = \sqrt{\frac{\omega_1^2}{\omega_1^2-m_1^2}} \frac{e^{i\phi_0}}{\sinh \omega_1(\tau-\tau_0)}, \hs{10} 
\varphi^j = 0~~~(j \not = 1),  
\label{eq:sol1_CPN}
\eeq
where $\omega_1 \equiv m_1 \sqrt{1+ N \epsilon g^2/m_1}$ 
and $(\tau_0, \phi_0)$ are moduli parameters. 
This is the single real bion solution, whose action is given by
\beq
S_{\rm rb}^1 = \frac{2\omega_1}{g^2} + N \epsilon 
\log \frac{\omega_1+m_1}{\omega_1-m_1}.
\label{eq:Srb1_CPN}
\eeq
The corresponding complex bion solution can be obtained by 
shifting $\tau_0 \rightarrow \tau_0 + \pi i / 2\omega_1$ 
and its action has an imaginary ambiguity related to the hidden 
topological angle  
\beq
S_{\rm cb}^1 = S_{\rm rb}^1 \pm N \epsilon \pi i ~~~~ 
\mbox{for $\arg \,g^{2} \rightarrow \pm 0$}. 
\label{eq:Scb1_CPN}
\eeq
Similarly, we can also embed the real and complex bion solution 
into $\varphi^i~(i=1,\cdots,N-1)$: 
\beq
\varphi^i = \sqrt{\frac{\omega_i^2}{\omega_i^2 - m_i^2}} 
\frac{e^{i\phi_0}}{\sinh \omega_i (\tau-\tau_0)}, \hs{10} 
\varphi^j = 0 ~~~(j \not = i),
\label{eq:solN_CPN} 
\eeq
where $\omega_i \equiv m_i \sqrt{1+ N \epsilon g^2/m_i}$. 
Therefore, the $\C P^{N-1}$ QM has $(N-1)$ types of real and 
complex single bion solutions as shown in Fig.\,\ref{fig:single-bions}.
We next look into the semiclassical contributions from these solutions.

\subsubsection{Semi-classical contribution of bion saddle points}
Next, let us calculate nonperturbative contributions to the ground state energy 
from the viewpoint of the saddle point method. 
We first note that the bion configuration can be rewritten into the kink-antikink form 
\beq
\varphi^i_{k \bar k} 
= \left( e^{\omega_i (\tau-\tau_+) - i \phi_+} - e^{-\omega_i(\tau-\tau_-)-i \phi_-}\right)^{-1}
~\mbox{with ~$e^{\pm \omega_i (\tau_\pm-\tau_0) +i (\phi_\pm - \phi_0)} 
= \sqrt{\frac{4\omega_i^2}{\omega_i^2-m_i^2}}$},
\label{eq:kink-antikink}
\eeq
In the weak coupling limit $g \rightarrow 0$, the kink and 
antikink are well separated ($|\tau_+ - \tau_-| \approx 
\log (4 m_i/N \epsilon g^2)$),
and the binding force between them becomes small. 
For such a configuration, 
the relative position $\tau_r=\tau_+-\tau_-$ and phase $\phi_r=\phi_+-\phi_-$ 
can be regarded as quasi-moduli parameters 
corresponding to the nearly flat directions around the saddle 
point configuration. 

As discussed in Appendix \ref{appendix:quasi_moduli}, 
we can decompose the degrees of freedom into 
the quasi-moduli parameters and orthogonal massive modes 
$\delta \varphi^j$: 
\beq
S ~=~ V_{\rm eff} + \sum_{j=1}^{N-1} \int d\tau \, 
\delta \bar \varphi^{\bar j} \Delta_j \delta \varphi^j + \mathcal O(g^4), 
\hs{10} \Delta_j = - \p_\tau^2 + {\mathcal V}_{i,j}(\tau).
\label{eq:SVeff_CPN}
\eeq
where $V_{\rm eff}$ is the effective potential between the 
well-separated kink and antikink and $\Delta_j$ is the 
differential operator with the potential ${\mathcal V}_{i,j}(\tau)$ 
for the $j$-th fluctuation $\delta \varphi^j$ 
on the $i$-th bion background
(see Appendix \ref{appendix:effpot} for details)
\beq
V_{\rm eff} &=& \frac{2m_i}{g^2} - \frac{4m_i}{g^2} \cos \phi_r e^{-m_i \tau_r} + N m_i \epsilon \tau_r. 
\label{eq:Veff_CPN}
\eeq
We can easily check that $V_{\rm eff}$ has the correct saddle points corresponding 
to the real and complex bion solutions \eqref{eq:kink-antikink}. 
The nonperturbative contribution from each single bion saddle point 
can be obtained from the quasi-moduli integral 
\beq
\lim_{\beta \rightarrow \infty} \left( - \frac{1}{\beta} \frac{Z_1}{Z_0} \right)_{i, {\rm bion}} &=& 
- \frac{8 m_i^4}{\pi g^2} \int d\tau_r d\phi_r \left( \prod_{j=1,\,j \not = i}^{N-1} \det \Delta_j^{-1} \right) \exp \left( - V_{\rm eff} \right). 
\label{eq:QMI_CPN}
\eeq
where $\det \Delta_j$ denotes the one-loop determinant for the fluctuation $\delta \varphi^j$. 
The overall factor $8 m_i^4/\pi g^2$, which includes the one-loop determinant for $\delta \varphi^i$, 
can be obtained in the same way as in the $\C P^1$ case \cite{Fujimori:2016ljw}.
As shown in Appendix \ref{appendix:one-loop}, 
the determinants $\det \Delta_j$ are given by
\beq
\prod_{j=1,\,j \not = i}^{N-1} \det \Delta_j^{-1} = A_i \, e^{(N-2)m_i \tau_r}. 
\label{eq:prodDet_CPN}
\eeq 
Hence the one-bion contribution can be rewritten as
\beq
- \lim_{\beta \rightarrow \infty} \frac{1}{\beta} \frac{Z_1}{Z_0} \ = \,  
- \sum_{i=1}^{N-1} \frac{8A_i m_i^4}{\pi g^2} \int d\tau_r d \phi_r \,  \exp \left( - V_{\rm eff}' \right), 
\label{eq:bionC_CPN}
 \eeq
with the modified effective potential 
\beq
V_{\rm eff}' = \frac{2m_i}{g^2} - \frac{4m_i}{g^2} \cos \phi_r \, e^{-m_i \tau_r} + 2 \epsilon' m_i \tau_r,
\label{eq:Veff2_CPN}
\eeq
where $\epsilon'$ is the ``renormalized" parameter related to $\epsilon$ as
\beq
\epsilon' = 1 - \frac{N}{2} (1-\epsilon). 
\label{eq:epprime_CPN}
\eeq
Since this modified kink-antikink potential is the same as that 
of the $\C P^1$ QM 
with $m \rightarrow m_i$ and $\epsilon \rightarrow \epsilon'$, 
the quasi-moduli integral can be performed as in the $\C P^1$ QM 
\cite{Fujimori:2016ljw}. 
Summing up all the single bion contributions, 
we obtain the following semiclassical contribution with an imaginary ambiguity: 
\beq
E_{1} \, = \, - \lim_{\beta \rightarrow \infty} \frac{1}{\beta} \frac{Z_1}{Z_0} 
\, = \, - \sum_{i=1}^{N-1} 2 m_i A_i \left( \frac{2m_i}{g^2} \right)^{2(1-\epsilon')} 
\frac{\Gamma(\epsilon')}{\Gamma(1-\epsilon')} e^{-\frac{2m_i}{g^2} \mp \pi i \epsilon'}.
\label{eq:E1_CPN}
\eeq

Let us consider the expansion around the SUSY point $\epsilon=1$. 
From Eq.\,(\ref{eq:E1_CPN}), 
we obtain the following first order expansion coefficient of the ground state energy:
\beq
E_1^{(1)} 
\, = \, \lim_{\epsilon \rightarrow 1} \p_\epsilon \left( - 
\lim_{\beta \rightarrow \infty} \frac{1}{\beta} \frac{Z_1}{Z_0} \right) 
\, = \, - \sum_{i=1}^{N-1} N m_i A_i e^{-\frac{2m_i}{g^2}}. 
\label{eq:E11bion_CPN}
\eeq
This precisely agrees with the leading order nonperturbative corrections 
which can be extracted from with the exact result \eqref{eq:CPN_E1}. 
The second order coefficient is given by
\beq
E^{(2)}_1 = \frac{1}{2} \lim_{\epsilon \rightarrow 1} \p_\epsilon^2 \left( - \lim_{\beta \rightarrow \infty} \frac{1}{\beta} \frac{Z_1}{Z_0} \right) 
= N^2 \sum_{i=1}^{N-1} m_i A_i e^{-\frac{2 m_i}{g^2}}\left[ \gamma + \log \frac{2m_i}{g^2} \pm \frac{\pi i}{2} + \mathcal O(g^2) \right],
\label{eq:E21bion_CPN}
\eeq
where $\gamma$ is the Euler-Mascheroni constant.
We here find the complete cancellation between the imaginary 
ambiguities of the $(N-1)$ types of the semiclassical single-bion 
contributions and those of the perturbative part 
\eqref{eq:CPN_E2_amb} arising from the $(N-1)$ Borel singularities.


\subsubsection{Multi-bion contributions}
Next, let us consider the multi-bion contributions to the partition function. 
Although we have not found multi-bion solutions except for the embedded ones, 
we assume that they consist of well-separated kink-antikink pairs 
each of which is either one of the $N-1$ types of bions. 
Then the semiclassical contribution to the partition function can be schematically written as
\beq
\frac{Z_p}{Z_0} \ = \ \sum_{i_1=1}^{N-1} \cdots \sum_{i_p=1}^{N-1} \, \int \prod_{n=1}^{2p} \Big[ d\tau_n d \phi_n \, B_n \exp \left( - V_{n,\,n-1} \right) \Big],  
\eeq
where $(\tau_n,\phi_n)$ are quasi-moduli parameters corresponding 
to the position and phase of the $n$-th constituent kink, 
$B_n$ are constants related to the integration measure and the one-loop determinant, 
and $(i_1,\cdots, i_p)$ denote the types of bions. 
Since the nearest-neighbor asymptotic interaction potential $V_{n,\,n-1}$ is 
a function of the relative quasi-moduli parameters
\beq
\tau_{n}^r \equiv \tau_{n} - \tau_{n-1},  \hs{10} 
\phi_{n}^r \equiv \phi_{n} - \phi_{n-1},  
\eeq 
the integral can be factorized by introducing the Lagrange multipliers $(\sigma, s)$ 
for the constraints
\beq
\sum_{n =1}^{2p} \tau_n^r = \beta, \hs{10} 
\sum_{n=1}^{2p} \phi_n^r = 0 + {\rm mod} \, 2 \pi,
\eeq
and changing the integration variables as $(\tau_n,\phi_n) \rightarrow (\tau_n^r,\phi_n^r)$.
Then, the summation over $(1,\cdots, i_p)$ can be recast into the following matrix form 
\beq
\frac{Z_p}{Z_0} \ = \ \frac{2\pi \beta}{p} \sum_{s=-\infty}^{\infty} \frac{1}{2\pi} \int \frac{d\sigma}{2\pi} \, 
e^{-i \beta \sigma} \, \tr ( \mathcal T ^p ),
\eeq
where the volume factor of the overall moduli $2\pi \beta$ is divided by 
$p$ since the bions are indistinguishable. 
The matrix $\mathcal T$ takes the form
\beq
(\mathcal T)_{ij} = \left( \frac{2m^2_i}{\pi g^2} \right)^2 A_i \, \mathcal I_i \mathcal J_{ij} \, e^{-\frac{2m_i}{g^2}},
\eeq
with 
\beq
\mathcal I_i = \int d \tau d \phi \, \exp \left( - V_i + i \sigma \tau + i s \phi \right), \hs{10}
\mathcal J_{ij} = \int d \tau d \phi \, \exp \left( - U_{ij} + i \sigma \tau + i s \phi \right),
\eeq
where $V_i$ is the interaction potential between kink and antikink of the same type
\beq
V_i = \frac{4m_i}{g^2} e^{-m_i \tau} \cos \phi + 2 m_i \epsilon' \tau, \hs{10} 
\epsilon' = 1 + \frac{N}{2}(\epsilon-1), 
\eeq
and $U_{ij}$ is that between a kink of $i$-th type and 
an antikink of $j$-th type. 
As in the $\C P^1$ case \cite{Fujimori:2017oab}, 
the leading contribution in the large-$\beta$ limit is given by
the residue at $\sigma = 0$ of the term with $s=0$
\beq
\frac{Z_p}{Z_0} \ \approx \ - \frac{i\beta}{p} \underset{\sigma = 0} {\rm Res} \left[ e^{-i \beta \sigma} \tr(\mathcal T^p) \right]_{s=0}.
\label{eq:bion_contribution}
\eeq
in the following, we focus only on the term with $s=0$. 

As in the single bion case, 
we can show by using the Lefschetz thimble method that
\beq
\mathcal I_i = \frac{\pi}{m_i} \left( \frac{2m_i}{g^2} e^{\pm \frac{\pi i}{2}} \right)^{i \frac{\sigma \ }{m_i} - 2 \epsilon'} 
\frac{\Gamma \left( \epsilon' - \frac{i\sigma \ }{2m_i} \right)}{\Gamma \left( 1 - \epsilon' + \frac{i\sigma \ }{2m_i} \right)}. 
\label{eq:I}
\eeq
Although the explicit form of $U_{ij}$ is not known, 
we can show from the fact that $U_{ij}$ vanishes for large $\tau$ that
$\mathcal J_{ij}$ has a pole at $\sigma=0$
\beq
\mathcal J_{ij} \ = \ \int d\tau d\phi \, \theta(\tau) e^{i \sigma \tau} + \int d\tau d\phi \left[ e^{-U_{ij}} - \theta(\tau) \right] e^{i \sigma \tau} 
\ = \ \frac{2\pi i}{\sigma} + \mathcal O(1).
\label{eq:J}
\eeq
From Eqs.\,\eqref{eq:I} and \eqref{eq:J}, 
we find that $\mathcal T$ and its derivative at $\epsilon=1$ can be expanded around $\sigma=0$ as 
\beq
(\mathcal T)_{ij} \Big|_{\epsilon=1} = A_i e^{-\frac{2m_i}{g^2}} + \mathcal O(\sigma), \hs{10}
\p_\epsilon (\mathcal T)_{ij} \Big|_{\epsilon=1} = \frac{i N m_i}{\sigma} A_i e^{-\frac{2m_i}{g^2}} + \mathcal O(1). 
\eeq
Since $\mathcal T$ has no pole at $\sigma = 0$ for $\epsilon=1$, 
the bion contribution to the partition function \eqref{eq:bion_contribution} vanishes
and hence there is no bion correction to the ground state energy 
\beq
\frac{Z_p}{Z_0} \bigg|_{\epsilon = 1} = 0 ~~~ \Longrightarrow ~~~ E_p^{(0)} = 0.
\eeq
Due to the pole of $\p_\epsilon (\mathcal T)_{ij} |_{\epsilon=1}$ at $\sigma=0$, 
there is nontrivial bion contribution to the first expansion coefficient 
\beq
E_p^{(1)} \, = - \lim_{\beta \rightarrow \infty} \frac{1}{\beta} \p_\epsilon \frac{Z_p}{Z_0} \bigg|_{\epsilon=1} 
\ = \ - N \left( \sum_{i=1}^{N-1} m_i A_i e^{-\frac{2m_i}{g^2}} \right) \left( \sum_{i=1}^{N-1} A_i e^{-\frac{2m_i}{g^2}} \right)^{p-1}.
\label{eq:E_p^1}
\eeq
Summing up the perturbative part and all the bion contributions, we obtain
\beq
E^{(1)} \, = \ E^{(1)}_0 + \sum_{p=1}^\infty E_p^{(1)} 
\ = \ \frac{N(N-1)}{2} g^2 - \sum_{i=1}^{N-1} m_i 
\left( 1 + \frac{N A_i e^{-\frac{2m_i}{g^2}}}{1 - \sum_{j=1}^{N-1} A_j e^{-\frac{2m_j}{g^2}}} \right).
\eeq 
Thus, the first expansion coefficient $E^{(1)}$ is completely reproduced by the semiclassical bion contributions. 

We note that as opposed to $E^{(1)}_p$, 
the $p$-bion contribution to the $n$-th expansion coefficient $E_p^{(n)}$ with $n \geq 2$ is an asymptotic series of $g^2$ 
\beq
E_p^{(n)} \ = \ E_{p,0}^{(n)} + E_{p,2}^{(n)} \, g^2 + E_{p,4}^{(n)} \, g^4 + \cdots 
\ = \ E_{p,0}^{(n)} + \int_0^\infty dt \, e^{-t/g^2} \tilde E_p^{(n)}(t).
\eeq
In the $\C P^1$ case \cite{Fujimori:2017oab}, 
the Borel transform $\tilde E_p^{(2)}(t)$ 
of the perturbative corrections to
the second order expansion coefficient 
around the $p$-bion background
has a pole which gives an imaginary ambiguity 
canceled by that of the semiclassical $p+1$-bion contribution
\beq
{\rm Im} \left( \int_0^\infty dt \, e^{-t/g^2} \tilde E_p^{(2)}(t) \right) \hs{5} 
\underset{\mathrm{cancellation}}{\leftarrow \hspace{-1.7mm}-\hspace{-1.7mm}-\hspace{-1.7mm}-\hspace{-1.7mm}-\hspace{-1.7mm}\rightarrow} 
\hs{5} {\rm Im} \, E_{p+1,0}^{(2)} \,. \hs{20}
\eeq
Although it is difficult to obtain the perturbative corrections directly from the $p$-bion background, 
we can determine $E^{(2)}$ by summing up all the semiclassical contributions
\beq
\sum_{p=1}^\infty E^{(2)}_{p,0} \ = \ m \frac{\cosh \frac{m}{g^2}}{\sinh^3 \! \frac{m}{g^2}} \left( \gamma + \log \frac{2m}{g^2} \pm \frac{\pi i}{2} \right), 
\eeq
using the dispersion relation, 
and imposing the symmetry $m \rightarrow - m$ as 
\footnote{The first term $g^2$ is added so that 
the perturbative part agrees with the Bender-Wu analysis. 
The argument here does not fix the degree of freedom to add 
any convergent series which is an even function of $m$.}
\beq
E^{(2)} = \, g^2 \! - 2 m \coth \frac{m}{g^2} \int_0^m \! \frac{d\mu}{\mu} \frac{\sinh^2 \! \frac{\mu}{g^2}}{\sinh^2 \! \frac{m}{g^2}} 
=\, g^2 \! - m \frac{\cosh \frac{m}{g^2}}{\sinh^3 \! \frac{m}{g^2}} \!
\left[ {\rm Chi} \bigg( \! \frac{2m}{g^2} \! \bigg) \! - \gamma - \log \frac{2m}{g^2} \right]. 
\eeq
Thus, taking advantage of the resurgence structure, we can 
completely reconstruct $E^{(2)}$ from the semiclassical 
bion contributions. 

In the $\C P^{N-1}$ case, 
the total semiclassical bion contributions can be formally written as
\beq
\sum_{p=1}^\infty E^{(2)}_{p,0} \, = \, 
N^2 \sum_{i=1}^{N-1} m_i A_i e^{-\frac{2m_i}{g^2}} 
\left[ \left( \gamma + \log \frac{2m_i}{g^2} \pm \frac{\pi i}{2} \right) Y_{ii} - \sum_{j=1}^{N-1} m_j A_j e^{-\frac{2m_j}{g^2}} Y_{ij} X_{ij} \right] \! , 
\label{eq:semiE2_CPN}
\eeq
where $X_{ij}$ and $Y_{ij}$ are defined by
\beq
X_{ij} \equiv \underset{\sigma=0}{\rm Res} \left( \frac{\mathcal J_{ij}}{2\pi \sigma} \right), \hs{5} 
Y_{ij} \equiv \frac{R_i R_j}{1-\sum_k A_k e^{-\frac{2m_k}{g^2}}}, \hs{5} 
R_i \equiv \frac{1 - \sum_k \frac{m_i-m_k}{m_i} A_k e^{-\frac{2m_k}{g^2}}}{1-\sum_k A_k e^{-\frac{2m_k}{g^2}}}.
\label{eq:X_Y}
\eeq
In the previous subsection, 
we have seen that 
the imaginary ambiguity of the single-bion contribution 
in Eq.\,\eqref{eq:semiE2_CPN} 
is canceled by that of the non-Borel summable perturbation series \eqref{eq:CPN_E2}.
Since the generating function $\langle 0 | 0 \rangle$ does not have a divergent asymptotic series, 
it is natural to expect that the higher bion sectors 
in $E^{(2)}= - \langle \Psi^{(1)}| H_{\epsilon=1} | \Psi^{(1)} \rangle/\langle 0 | 0 \rangle$ 
have the same cancellation structure: 
the imaginary ambiguity of the semiclassical $(p+1)$-bion contribution is 
canceled by that of the non-Borel summable perturbation series around the $p$-bion background. 
Thus we expect that it is also possible to recover $E^{(2)}$ 
from the semiclassical bion contributions \eqref{eq:semiE2_CPN}
in a parallel manner to the $\C P^{1}$ QM. 
It would be interesting to check if there is such resurgence 
structure in the $\C P^{N-1}$ QM by explicitly determining 
$E^{(2)}$ and $X_{ij}$ in Eqs.\,\eqref{eq:X_Y} and 
\eqref{eq:semiE2_CPN}. 


\subsection{Quasi-exact solvability of $\C P^{N-1}$ quantum mechanics}
As in the case of the sine-Gordon QM 
discussed in the previous section, 
the $\C P^{N-1}$ QM becomes quasi-exactly solvable 
at some specific points in the parameter space. 
By introducing a deformation parameter around those QES points, 
we obtain exact results for the expansion coefficients of 
the ground state energy, 
which show nontrivial resurgence structure around the QES points.

Here we focus on the sector with vanishing conserved charges, 
where wave functions are independent of $\arg \varphi^i$. 
For later convenience, 
we define the following new variables:
$x_i = \frac{|\varphi^i|^2}{1 + |\varphi^k|^2}$. 
Redefining the wave function as
\beq
\Psi = \Psi_0 u(x_i), \hs{10} 
\Psi_0 = \exp \left( - \frac{\mu}{g^2} \right), 
\label{eq:WF_QES_CPN}
\eeq
we can rewrite the Schr\"odinger equation as $\tilde H u = E u$ with
\beq
\tilde H = g^2 \sum_{i=1}^{N-1} 
\Bigg[ \left( T_i{}^N - \frac{2m_i}{g^2} \right) \left( T_i{}^i - T_N{}^i \right) + \frac{1-\epsilon}{2} \left( N (T_i{}^i -1) + \frac{2m_i}{g^2} \right) \Bigg],
\label{eq:H_QES_CPN}
\eeq
where $T_I{}^J$ are the following differential operators 
\beq
T_i{}^j = - x_j \frac{\p}{\p x_i}, \hs{5} T_i{}^N = \frac{\p}{\p x_i}, \hs{5} T_N{}^i = - x_i T_N{}^N, \hs{5} T_N{}^N = \sum_{i=1}^{N-1} x_i \frac{\p}{\p x_i} - \frac{N}{2}(\epsilon-1). 
\label{eq:T_QES_CPN}
\eeq
The operators $T_I{}^J~(I,J=1,\cdots,N)$ satisfy the $\mathfrak{gl}(N,\C)$ algebra 
\beq
[T_I{}^J, T_K{}^L] = \delta_K{}^J T_I{}^L - \delta_I{}^L T_K{}^J.
\eeq 
The quadratic Casimir invariant is given by
\beq
\sum_{I,J=1}^N T_I{}^J T_J{}^I =(\epsilon'-1) (\epsilon' + N - 2), \hs{10} \epsilon' = 1 + \frac{N}{2}(\epsilon-1).  
\label{eq:casimir_CPN}
\eeq
If $\epsilon'$ is an integer, 
the action of the operators $T_I{}^J$
is closed on the set of polynomials of $x_i$ of degree $\epsilon'-1$. 
Therefore, we can find eigenfunctions of $\tilde H$ 
by using the polynomial ansatz for $u(x_i)$
(corresponding to the symmetric representation of $\mathfrak{gl}(N,\C)$ with $\epsilon'-1$ indices).

Since $\epsilon'=1$ is equivalent to $\epsilon =1$, 
the deformation from $\epsilon'=1$ is nothing but the SUSY-breaking deformation 
whose resurgence structure has been already discussed in the previous three subsections.
Now, we consider the case of $\epsilon' = 2$.
We can find eigenfunctions by using the ansatz $u = a_N + \sum_{i=1}^{N-1} a_i x_i$. 
Substituting into the Schr\"odinger equation, we find that
\beq
u = 1 + \sum_{i=1}^{N-1} \frac{m_i x_i}{M-m_i}, \hs{10} 
E = g^2 N + \frac{2}{N} \left( M + \sum_{i=1}^{N-1} (M-m_i) \right), 
\label{eq:ep2_CPN}
\eeq
where $M$ is one of the solutions of the equation
\beq
\sum_{i=1}^{N-1} \frac{1}{m_i-M} - \frac{1}{M} = \frac{2}{g^2}.
\eeq
There are $N$ solutions corresponding 
to the $N$-dimensional fundamental representation of $\mathfrak{gl}(N,\C)$. 
Let us consider the small $\epsilon$ expansion 
of the smallest eigenvalue corresponding to the ground state energy 
\beq
E^{(1)} = \lim_{\epsilon' \rightarrow 2} \frac{\p}{\p \epsilon} E 
= \frac{\langle \Psi | \delta H |\Psi \rangle}{\langle \Psi | \Psi \rangle}. 
\label{eq:E1_ep2_CPN}
\eeq
Here we again note that $\delta H = \p_\epsilon H$ is given by 
$\delta H =-\Delta\mu = - \sum_{i=1}^{N-1} m_i (1 - N x_i)$.
Evaluating the integral in Eq.\,\eqref{eq:E1_ep2_CPN}, we obtain 
\beq
E^{(1)} = - \sum_{i=1}^{N-1} m_i \left[ (1 + \mathcal O(g^2)) + \frac{\displaystyle N \tilde A_i e^{-\frac{2m_i}{g^2}}}{1 - \sum_i \tilde A_i e^{-\frac{2m_i}{g^2}}} \right], 
\label{eq:E1_ep2_CPN2}
\eeq
where $\tilde A_i$ are constants which have the following weak coupling forms
\beq
\tilde A_i = \left( \frac{g^2}{2m_i} \right)^{2} \left[ A_i + \mathcal O(g^2) \right] 
= \left( \frac{g^2}{2m_i} \right)^{2} \left[ \prod_{i \not = j} \frac{m_j}{m_j-m_i} + \mathcal O(g^2) \right].
\label{eq:Ai_ep2_CPN}
\eeq
Thus we find that the $p$-th order nonperturbative correction takes the form
\beq
E^{(1)}_{p} = - N \left( \sum_{i=1}^{N-1} m_i \tilde A_i e^{-\frac{2m_i}{g^2}} \right)
\left( \sum_{i=1}^{N-1} \tilde A_i e^{-\frac{2m_i}{g^2}} \right)^{p-1}.
\label{eq:E1_ep2_CPN_final} 
\eeq
We can check the agreement between 
the leading order part of Eq.\,\eqref{eq:E1_ep2_CPN_final} for small $g$ and 
the semiclassical multi-bion contribution (\ref{eq:bion_contribution}) expanded around $\epsilon'=2$.
Thus, in the weak coupling limit, 
the nonperturbative corrections in the exact result are 
correctly reproduced by semiclassical multi-bion contributions 
not only around the SUSY regime 
but also around the near-QES regime of the ${\mathbb C}P^{N-1}$ QM.


\section{Resurgence Structure in Squashed $\C P^1$ quantum mechanics}
\label{sec:sCPN}

In this section, we briefly discuss another type of model
belonging to the class described by chiral multiplets 
in which the ${\mathcal O}(\delta \epsilon)$ ground state energy has nontrivial resurgence structure.

\begin{figure}[t]
\includegraphics[width=140mm, viewport = 14 111 531 280]{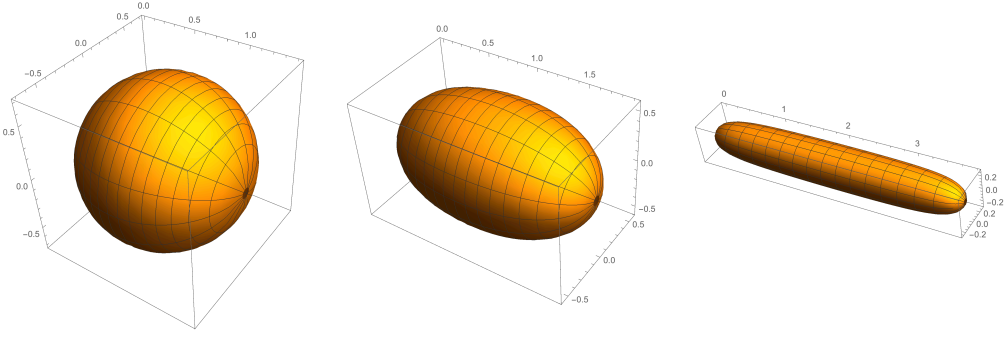}
\caption{The squashed $\C P^1$ with $a=0$ (left), $a=1$ (center), 
$a=10$ (right).}
\label{fig:deformedcp_1}
\end{figure}

We here focus on the model described by the K\"ahler potential 
with a 
parameter $a\ge0$ 
\beq
K = \log \left({1\over{1-x}}\right) + a x^{2} \,,
\label{eq:defCP1_Kp}
\eeq
where $x$ is the function of $\varphi$ determined by
\beq
|\varphi| = e^{ax} \sqrt{\frac{x}{1-x}}.
\eeq 
The target space is the squashed $\C P^{1}$ (see Fig.\,\ref{fig:deformedcp_1}) whose metric is given by
\beq
ds^2 = \p_\varphi \p_{\bar \varphi} K d \varphi d \bar \varphi = \frac{1}{2}( U dx^2 + U^{-1} d \arg \varphi^2), \hs{5}
U \equiv a + \frac{1}{2} \left( {1\over{x}}+{1\over{1-x}} \right).
\eeq
This model reduces to the standard $\C P^{1}$ QM for $a=0$.

Now let us consider the small $\delta \epsilon = \epsilon - 1$ 
expansion of the ground state energy
\beq
E^{(1)} = \frac{\langle 0 | \delta H | 0 \rangle}
{\langle 0 | 0 \rangle}, \hs{10}
\langle \varphi | 0 \rangle = \exp \left( - \frac{\mu}{g^2} \right), 
\hs{10} \mu=mx. 
\eeq
It is quite notable that the perturbation Hamiltonian 
$\delta H =-\Delta\mu = -G^{-1}\p{\bar \p}\mu$ is not a polynomial 
but the following rational function
\beq
\delta H \ = \, - {m\over{2}} \partial_{x} U^{-1} \ 
= \ m \frac{2x-1}{(1+2a(1-x)x)^2}.
\eeq
This implies that the trans-series expression for the first 
expansion coefficient $E^{(1)}$ has non-Borel-summable $g^2$ 
series at each order of the nonperturbative exponential. 
This is a crucial difference from the standard $\C P^{N-1}$ QM.

Again the generating function $\langle 0|0\rangle$ can be 
calculated by the localization formula. 
It does not depend on $a$ and hence it is identical to that 
in the standard $\C P^1$ QM 
\beq
\langle 0|0\rangle = {\pi g^{2}\over{2m}}  
\left(1-e^{- {2m\over{g^{2}}}}\right).
\label{eq:00_defCP1}
\eeq
This is finite-order in terms of $g^2$ 
at each order of nonperturbative exponential. 
Evaluating the integral
\beq
\langle 0|\delta H|0\rangle = 
- \frac{\pi m}{2} \int_0^1 dx \, e^{-\frac{2\mu}{g^2}} \p_x U^{-1} \,,
\label{eq:0dH0_defCP1}
\eeq
we obtain the first order coefficient as 
\beq
E^{(1)} = -{m^{2}\over{ag^{2}}}\left(
1+{2m\over{g^{2}}}{X\over{1-e^{-{2m\over{g^{2}}}}}}
\right), \hs{10} 
X=\int_{0}^{1} dx \frac{e^{-\frac{2mx}{g^2}}}{1+2ax(1-x)} \,.
\label{eq:E1_defCP1}
\eeq 
The quantity $X$ is a linear combination of the exponential 
integral function whose asymptotic series in powers of $g^2$ 
can be obtained by changing variable to $t=2mx/g^2$ and 
taking the upper limit of integration $2m/g^2\to\infty$ 
\begin{eqnarray}
X&=&
\frac{1}{2a(2b-1)}
\int_0^{2m/g^2}dt \, e^{-t}\left(
\frac{1}{t+\frac{2m(b-1)}{g^2}}-\frac{1}{t-\frac{2mb}{g^2}}\right), 
\nonumber \\
&\approx &
\frac{1}{2a(2b-1)}
\left[
\sum_{n=0}^\infty n!\left(\frac{-g^2}{2m(b-1)}\right)^{n+1}
+
\sum_{n=0}^{\infty} n! \left(\frac{g^2}{2mb}\right)^{n+1}
\right], 
\end{eqnarray}
where $b = {1\over{2}}+ \sqrt{a+2\over{4a}}$. 
Since the perturbation series of $X$ is a sum of Borel 
summable and non-summable series, the Borel resummation 
of the perturbation series gives the perturbative contribution 
$X_{\rm pert}$ containing the imaginary ambiguities as 
\begin{equation}
X_{\rm pert}=\frac{1}{2a(2b-1)}f\left(\frac{2m}{g^2}\mp i0\right), 
\hs{5}
f(z)=
\int_0^{\infty}dt \, e^{-t}\left(\frac{1}{t+(b-1)z}-
\frac{1}{t-bz}\right). 
\end{equation}
The remaining part of $X$ is multiplied by a single 
power of the nonperturbative exponential 
\begin{equation}
X-X_{\rm pert}=\frac{e^{-\frac{2m}{g^2}}}{2a(2b-1)}
g\left(\frac{2m}{g^2}\mp i0\right), 
\hs{5}
g(z)=
\int_0^{\infty}dt \, e^{-t}\left(-\frac{1}{t+bz}+
\frac{1}{t-(b-1)z}\right), 
\end{equation}
which corresponds to the Borel resummation of a sum of Borel 
summable and non-summable asymptotic series $g^2$ on the single 
bion background. 
Expanding with respect to $e^{-\frac{2m}{g^2}}$, 
we obtain the trans-series expression 
$E^{(1)}=\sum_{p=0}^{\infty} E^{(1)}_{p}$ with 
\beq
E_{0}^{(1)} &=& -\frac{m^2}{a g^2} 
-\frac{m^3}{a^2g^4(2b-1)} 
f\left(\frac{2m}{g^2}\mp i0\right) , \\
E_{p}^{(1)} &=& -{m^{3}\over{a^2 g^{4}(2b-1)}}
\left[f\left(\frac{2m}{g^2}\mp i0\right)
+g\left(\frac{2m}{g^2}\mp i0\right)\right] 
e^{-\frac{2pm}{g^2}} \,.
\label{eq:E0Ep_defCP1}
\eeq
Here, the $\pm$ signs correspond to 
${\rm arg}\,g^{2} \rightarrow \pm0$.
At each order of the non-perturbative exponential 
$e^{-\frac{2m}{g^2}}$, 
$E_{p}^{(1)}$ has the following imaginary ambiguities 
due to the poles at $t = 2mb/g^2$ and $t=2m(b-1)/g^2$: 
\beq
&&{\rm Im} \, E_0^{(1)} \, = \, \pm \frac{\pi m^3}{a^2 g^4} 
\frac{1}{2b-1} e^{-\frac{2b m}{g^2}}.
\\
&&{\rm Im} \, E_p^{(1)} \, = \, \pm \frac{\pi m^3}{a^2 g^4} 
\frac{1}{2b-1} \left[ e^{-\frac{2(p+b)m}{g^2}} 
- e^{-\frac{2(p-1+b)m}{g^2}}\right] \,\,\,\,\,\,\,\,\,\,\,\,\,(p\geq 1).  
\eeq
These imaginary ambiguities cancel out between the adjacent sectors.
Thus, we conclude that the resurgence structure 
of the ${\mathcal O}(\delta\epsilon)$ ground state energy is nontrivial.  
Nevertheless, it has a simple structure where imaginary 
ambiguities cancel between the adjacent sectors of the $p$ 
and $p+1$ power of the nonperturbative exponentials. 
It is worth noting that the cancellation mechanism of the 
imaginary ambiguities seems different from that in the standard 
$\C P^1$ QM. 
For example, the ambiguities have non-integer powers of 
$e^{-\frac{2m}{g^2}}$ and all of them are from the Borel 
resummation of the asymptotic series.
It is interesting to interpret these cancellation mechanism 
from the viewpoint of the complexified path integral.

\section{Summary and Discussion}
\label{sec:SD}

In this paper, we investigated the resurgence structure 
in SUSY quantum mechanics
with emphasis on the expansion around 
the SUSY and quasi-exactly solvable (QES) parameter regimes. 
First, we generically showed that 
bions play a vital role in nonperturbative contributions 
based on the Lefschetz-thimble decomposition.
We discussed two different classes of SUSY models: 
(i) quantum mechanics on a Riemannian manifold 
described by real multiplets 
and (ii) quantum mechanics on a K\"ahler manifold 
described by chiral multiplets. 

In the models belonging to class (i), 
the generating function $\langle 0|0\rangle$ 
has a non-Borel summable asymptotic series 
which gives rise to an imaginary ambiguity 
at each order of nonperturbative exponentials in trans-series. 
This property of the generating function $\langle 0|0\rangle$ 
provides the ${\mathcal O}(\delta\epsilon)$ ground state energy
with rich resurgence structure. 
As an example of a model in this class, 
we discussed the sine-Gordon model in Sec.\,\ref{sec:SG}. 
Using the Rayleigh-Schr\"odinger perturbation theory, we obtained 
the exact ${\mathcal O}(\delta\epsilon)$ ground state energy, 
which was expressed as a trans-series of non-Borel-summable series 
with nonperturbative exponentials corresponding to multi-bions. 
We showed that the semiclassical contributions 
from the complex multi-bion solutions are 
in agreement with those in the exact result
including the imaginary ambiguities which 
cancel those in the other sectors.

In the models belonging to class (ii),  
the generating function $\langle 0|0\rangle$ 
can be exactly calculated by 
the Duistermaat-Heckman localization formula.
Since it has a convergent (finite) power series of $g^{2}$ 
in each sector of non-perturbative exponentials,
the ${\mathcal O}(\delta\epsilon)$ ground state energy 
has relatively simple resurgence structure. 
As an example of a model in this class, 
the $\C P^{N-1}$ QM was discussed in Sec.\,\ref{sec:CPN}.
We determined the exact 
${\mathcal O}(\delta\epsilon)$ ground state energy 
and showed that it has trivial resurgence structure 
with no imaginary ambiguities in the trans-series. 
This property enabled us to completely reconstruct 
the ${\mathcal O}(\delta\epsilon)$ ground state energy 
from the semiclassical multi-bion contributions. 
The ground state energy has non-trivial resurgence structure 
at ${\mathcal O}(\delta\epsilon^{2})$ and higher orders. 
We found $N-1$ types of real and complex bion solutions 
and showed the resurgence structure 
with the cancellation between imaginary ambiguities 
arising from the Borel resummation of the perturbation series 
around the perturbative vacuum (zero-bion background) 
and the semiclassical contributions of the single-bion solutions. 
As shown in the example of the squashed $\C P^1$ QM, 
the ${\mathcal O}(\delta\epsilon)$ ground state energy of 
a generic model in class (ii) has richer resurgence structure 
with a nontrivial cancellation of imaginary ambiguities in trans-series.

This work reveals that a broad class of 
quantum mechanical models have a nontrivial resurgent structure, 
where exact results for physical quantities are 
expressed as resurgent trans-series 
which consist of perturbative Borel resummations 
and complex multi-bion contributions.
The cancellation of imaginary ambiguities enables us 
to reproduce a contribution of one sector from another 
by use of the dispersion relation of Cauchy's theorem.
However, in some special cases such as 
the ${\mathcal O}(\delta\epsilon)$ ground state energy 
in the $\C P^{N-1}$ QM, 
each order term of nonperturbative exponentials 
produces no imaginary ambiguities 
and the resurgence structure is trivial.
This situation is similar to the case of the partition function 
in the ${\mathcal N}=2$ SYM \cite{Aniceto:2014hoa, Honda:2016mvg} obtained by the localization method, 
where each sector of the trans-series does not talk to other sector. 
One of our next plans is 
to apply our analysis to solvable field theoretical models 
where the localization method is applicable 
to find nontrivial resurgent structure 
in the expansion with respect to deformation parameters.
It has been observed that the information on the level number 
of the perturbation series on zero-bion background gives 
all $p$-bions contributions \cite{Dunne:2016jsr}. 
It is an interesting future task to obtain resurgent trans-series 
for states other than the ground state.

One of the goals is to discuss 
roles of complexified solutions in Yang-Mills or QCD in four dimensions. 
In 1970's and 80's, complex instanton solutions were 
discussed in gauge theories with complexified gauge groups \cite{Dolan:1977hs,Burns:1983us}. 
It will be interesting to discuss contributions 
from these complex solutions in terms of resurgence theory.
Another way of studying complexified solutions in Yang-Mills theory is 
to consider $U(N)$ Yang-Mills theory coupled with Higgs fields 
in the fundamental representation.
In the Higgs phase, there exists a non-Abelian vortex whose 
low-energy dynamics is effectively described by 
the ${\mathbb C}P^{N-1}$ model localized around the vortex 
\cite{Hanany:2003hp,Auzzi:2003fs,Eto:2005yh,Eto:2006cx}. 
By introducing appropriate fermions coupled 
in the original bulk theory, 
we can localize fermion quasi-zero modes around the vortex and 
the ${\mathbb C}P^{N-1}$ model is coupled to the fermions.
For the case of the SUSY bulk theory, 
the vortex can be BPS and 
the SUSY ${\mathbb C}P^{N-1}$ model is obtained 
as the vortex theory.
It means that our complexified solutions are 
able to be embedded into it. 
${\mathbb C}P^{N-1}$ model instantons in the vortex theory 
correspond to Yang-Mills instantons 
in the bulk theory \cite{Eto:2004rz}. 
Therefore, complexified bions can be interpreted 
as those in (complexified) Yang-Mills theory in the bulk. 
A decoupling limit of the Higgs phase leads to isolation of 
complexified solutions in Yang-Mills theory.
We hope that these are useful to reveal 
the resurgence structure of Yang-Mills theory or 
QCD in four dimensions.



\begin{acknowledgments}
The authors are grateful to the organizers and participants 
of ``Resurgence in Gauge and String Theories 2016" at IST, 
Lisbon and ``Resurgence at Kavli IPMU" at IPMU, U. of Tokyo 
for giving them a chance to deepen their ideas. 
This work is supported by 
the Ministry of Education, Culture, Sports, Science,
and Technology(MEXT)-Supported Program for the Strategic Research Foundation
at Private Universities ``Topological Science" (Grant No. S1511006).
This work is also supported in part 
by  the Japan Society for the 
Promotion of Science (JSPS) 
Grant-in-Aid for Scientific Research
(KAKENHI) Grant Numbers 
(16K17677 (T.\ M.), 16H03984 (M.\ N.) and 
25400241 (N.\ S.)).
The work of M.N. is also supported in part 
by a Grant-in-Aid for Scientific Research on Innovative Areas
``Topological Materials Science"
(KAKENHI Grant No. 15H05855) 
from MEXT of Japan.
\end{acknowledgments}


\appendix

\section{Supersymmetric quantum mechanics}
\label{appendix:SQM}
In this appendix, we review the models of SUSY QM discussed 
in this paper. Consider the SUSY algebra 
\beq
\{ Q , \bar Q \} = H - P,
\eeq
where $H$ is the Hamiltonian and 
$P$ denotes the central charge, 
which may exist when there is an internal symmetry which 
commutes with $Q$ and $\bar Q$. 
We consider two types of supermultiplets, namely real and chiral multiplets, 
each of which has a bosonic degree of freedom in its lowest component.

\subsection{Real multiplets}\label{subsec:real}
Let us first consider the SUSY QM described by real multiplets. 
For simplicity, we assume that there is no central charge 
appearing in the superalgebra ($P=0$). 
The SUSY transformation of a real multiplet is given by
\begin{align}
\delta \varphi &= \frac{1}{2} \left( \varepsilon \psi + \bar \varepsilon \bar \psi \right), 
& 
\delta \psi &= \bar \varepsilon ( i \p_t \varphi + F ), \\
\hs{10}
\delta F &= \frac{i}{2}( \varepsilon \p_t \psi - \bar \varepsilon \p_t \bar \psi), 
&
\delta \bar \psi & = \varepsilon ( i \p_t \varphi - F ),
\hs{20}
\end{align}
where $\varepsilon$ and $\bar \varepsilon$ are transformation parameters for $Q$ and $\bar Q$. 
Let us consider a Riemaniann manifold $\mathcal M$ parametrized by the scalar components $\varphi^i$. 
After integrating out the auxiliary fields $F^i$, 
the SUSY Lagrangian takes the form
\beq
L &=& \frac{1}{g^2} \left[ \frac{1}{4} G_{ij} \left( \dot \varphi^i \dot \varphi^j + i \bar \psi^i \D_t \psi^j \right) - \frac{1}{32} R_{ijkl} \psi^i \psi^j \bar \psi^k \bar \psi^l - G^{ij} \p_i W \p_j W + \frac{1}{2} \nabla_i \p_j W \psi^i \bar \psi^j \right],
\eeq
where $G_{ij}$ is the Riemaniann metric, 
$\D_t \psi^i = \p_t \psi^i + \Gamma^i_{jk} \dot \varphi^j \psi^j $, 
$\Gamma^i_{ij}$ is the Christoffel symbol, 
$R_{ijkl}$ is the Riemannian curvature tensor and 
$W$ is the superpotential, which can be an arbitrary function on $\mathcal M$.

\paragraph{BPS kink solution\\}
Next let us discuss BPS kink configurations in this model 
(see, e.g. \cite{Dvali:1996bg,Chibisov:1997rc,Isozumi:2004jc,Isozumi:2004va} for BPS kinks). 
After the Wick rotation, 
the bosonic part of the Euclidean action can be rewritten as
\beq
S_E = \frac{1}{g^2} \int d\tau \, G_{ij} \left( \frac{1}{2} \p_\tau \varphi^i \pm G^{ik} \p_k W \right) \left( \frac{1}{2} \p_\tau \varphi^j \pm G^{jl} \p_l W \right) \mp \frac{1}{g^2} \int d W.
\eeq
This form of the Euclidean action implies that 
there exist BPS kink solutions obeying the flow equation
\beq
\frac{1}{2} \p_t \varphi^i = \mp G^{ik} \p_k W. 
\eeq
Such BPS kinks correspond to tunneling processes
between two minima of the potential, i.e. saddle points of $W$. 
For a BPS kink interpolating two saddle points $s$ and $s'$, 
the on-shell value of the Euclidean action is given by
\beq
S_{E,(s,s')} = \left| \frac{W_s - W_{s'}}{g^2} \right|, 
\eeq
where $W_s$ and $W_{s'}$ are the values of the superpotential 
at the corresponding saddle points.
Typical nonperturbative effects for the ground state are of order $\exp\left(-2S_{E,(s,s')} \right)$, 
which implies that they are given by bound states of kink and antikink, i.e. bion configurations. 

\paragraph{Hamiltonian and ground state wave function \\}
Let us quantize the system by introducing the commutation relation between the canonical coordinates, 
whose nontrivial part is given by
\beq
[\varphi^i, p_j] = i \delta^i{}_j, \hs{10} 
\{ \psi^i, \pi_{\psi^j} \} = i \delta^i{}_j,
\eeq
where $p_i$ and $\pi_{\psi^i}$ are the conjugate momenta 
\beq
p_i = \frac{\p L}{\p \dot \varphi^i} = \frac{1}{2g^2} G_{ij} \left[ \dot \varphi^j + \frac{i}{2} \Gamma^j_{kl} \bar \psi^k \psi^l \right], \hs{10}
\pi_{\psi^i} = \frac{\p L}{\p \dot \psi^i} = \frac{i}{4g^2} G_{ij} \bar \psi^j. \hs{10}
\eeq
Let us project the Hilbert space onto the subspace with the lowest fermion number $F = G_{ij} \psi^i \bar \psi^j$:
\beq
0 = \langle \Psi | F | \Psi \rangle = \sum_{a=1}^n || e^a{}_i \bar \psi^i |\Psi \rangle ||^2 ~~~\Longrightarrow~~~ \bar \psi^i | \Psi \rangle = 0,
\eeq
where $e^a{}_i$ are the vielbein defined by $G_{ij} = \sum_{a=1}^n e^a{}_i e^a{}_j$. 
On this subspace, the Hamiltonian $H = \{ Q, \bar Q \}$ reduces to
\beq
H |\Psi \rangle = \bar Q Q | \Psi \rangle ~~~\Longleftrightarrow~~~ 
- g^2 G^{ij} \left( \nabla_i - \frac{1}{g^2} \p_i W \right) \left( \p_j + \frac{1}{g^2} \p_j W \right) \langle \varphi | \Psi \rangle.
\eeq
where we have used the following explicit form of the supercharges written in terms of the canonical coordinate:
\beq
Q = \frac{1}{2} \psi^i \left( p_i - \frac{i}{g^2} \p_i W \right), \hs{10} 
\bar Q = \frac{1}{2} \bar \psi^i \left( p_i - \frac{i}{4g^2} G_{il} \Gamma^l_{jk} \psi^k \bar \psi^j + \frac{i}{g^2} \p_i W \right). 
\eeq
Therefore, the SUSY ground state, 
which has the lowest energy $H|\Psi \rangle = 0$, 
is the one annihilated by $p_i - \frac{i}{g^2} \p_i W$, i.e. 
\beq
\langle \varphi | \Psi \rangle = \exp \left( - \frac{W}{g^2} \right).
\eeq

\subsection{Chiral multiplets}
Next let us consider the SUSY QM described by chiral multiplets. 
Such a model can be obtained from the corresponding 2d nonlinear sigma model
with $\mathcal N = (2,0)$ SUSY
\beq
\{Q, \bar Q \} = H - P,
\eeq
where $P$ is the spatial momentum. 
Introducing the Grassmannian coordinates $\theta$ and $\bar \theta$, 
we can associate the supercharges $Q$ and $\bar Q$ with the differential operators
\beq
Q = \frac{\p}{\p\theta} + i \bar \theta \p_-, \hs{10} \bar Q = - \frac{\p}{\p \bar \theta} - i \theta \p_-, 
\eeq
with $\p_\pm = \frac{1}{2}(\p_t \pm \p_x)$. 
The chiral and anti-chiral superfields are respectively defined as 
those annihilated by the differential operators,
\beq
D = \frac{\p}{\p \theta} - i \bar \theta \p_+, \hs{10}
\bar D = - \frac{\p}{\p \bar \theta} + i \theta \p_+, 
\eeq
which anti-commute with $Q$ and $\bar Q$. 
The explicit forms of the chiral and anti-chiral superfields are given by
\beq
\bar D \Phi = 0 ~~~&\Longrightarrow&~~~ \Phi = \varphi + \theta \psi + i \bar \theta \theta \p_+ \varphi, \\
D \bar \Phi = 0 ~~~&\Longrightarrow&~~~ \bar \Phi = \bar \varphi - \bar \theta \bar \psi - i \bar \theta \theta \p_+ \bar \varphi. 
\eeq 
The Lagrangian of the nonlinear sigma model 
with K\"ahler potential $K$ and K\"ahler metric $G_{i \bar j} = \p_i \bar \p_{\bar j} K$ can be written as
\beq
\mathcal L \ = \ \frac{1}{g_{2d}^2} \int d\theta d \bar \theta \, 2 i \p_- \Phi^i \p_i K
\ = \ \frac{1}{g_{2d}^2} G_{i \bar j} \left(- \p_\mu \varphi^i \p^\mu \bar \varphi^{\bar j} + 2 i \bar \psi^{\bar j} \D_- \psi^i \right) + \cdots,
\label{eq:L_2d}
\eeq
where $\D_- \psi^i = \p_- \psi^i + \Gamma^i_{jk} \p_- \varphi^j \psi^k$ and 
$\cdots$ denotes total derivative terms. 

Let us assume that the target manifold has a holomorphic isometry with moment map $\mu$. 
The corresponding holomorphic Killing vector $\Xi \equiv \xi^i \p_i + \bar \xi^i \bar \p_i$ satisfies 
\beq
\xi^i = i G^{\bar j i} \bar \p_j \mu, \hs{10}
\bar \xi^{\bar i} = - i G^{\bar j i} \p_i \mu,
\hs{10} 
\bar \p_{\bar j} \xi^i = \p_i \bar \xi^{\bar j} = 0.
\eeq
The SUSY QM of chiral multiplets discussed in this paper
can be obtained from Eq.\,\eqref{eq:L_2d} by dimensional reduction twisted by the isometry
\beq
L = \frac{1}{g^2}  G_{i \bar j} 
\left[ \dot \varphi^i \dot{\bar \varphi}^{\bar j} - \xi^i \bar \xi^{\bar j} + i \bar \psi^{\bar j} \D_t \psi^i - i \nabla_k \xi^i \bar \psi^{\bar j} \psi^k \right],
\label{eq:KahlerL1}
\eeq
where $\D_t \psi^i = \dot \psi^i + \Gamma^i_{jk} \dot \varphi^j \psi^k$, 
$\nabla_k \xi^i = \p_k \xi^i + \Gamma^i_{jk} \xi^j$,
$\Gamma_{jk}^i = \p_j G_{k \bar l} G^{\bar l i}$ and 
\beq
\frac{1}{g^2} = \frac{1}{g_{2d}^2} \times 2\pi \{\mbox{compactification radius}\}.
\eeq 

\paragraph{Symmetry \\}
The SUSY transformations for the components of a chiral multiplet are given by
\beq
\delta \varphi^i = \varepsilon \psi^i, &\hs{10}& \delta \psi^i = i \bar \varepsilon (\dot \varphi^i + \xi^i), \\
\delta \bar \varphi^i{\bar i}= \bar \varepsilon \bar \psi^{\bar i}, &\hs{10}& \delta \bar \psi^{\bar i} = i \varepsilon (\dot{\bar \varphi}^{\bar i} + \bar \xi^{\bar i}),
\eeq
where $\varepsilon$ and $\bar \varepsilon$ are transformation parameters for $Q$ and $\bar Q$. 
We can see from these SUSY transformations that the SUSY
algebra takes the form 
\beq
\{ Q , \bar Q \} = H + P,
\eeq
where $P$ is given by the Noether charge $q$ 
for the holomorphic isometry 
$\delta_q \varphi^i = \xi^i,  \delta_q \psi^i = \psi^j \p_j \xi^i$: 
\beq
q = \frac{1}{g^2} G_{i \bar j} \left( \xi^i \dot{\bar \varphi}^{\bar j} + \dot \varphi^i \bar \xi^{\bar j} + i \bar \psi^{\bar j} \psi^k \nabla_k \xi^i \right).
\label{eq:Ncharge1}
\eeq

\paragraph{BPS kink \\}
The original 2d system \eqref{eq:L_2d} in the Euclidean spacetime 
has instanton solutions characterized by the topological charge \cite{Polyakov:1975yp}
\begin{eqnarray}
Q = \frac{1}{2\pi} \int dx_1 dx_2 \, i \epsilon^{\mu\nu} G_{i \bar j} \p_\mu \varphi^i \p_\nu \bar \varphi^{\bar j} \,.
\label{eq:2dtopcharge}
\end{eqnarray}
After the twisted dimensional reduction, 
such an instanton decomposes into fractional instantons
characterized by the topological charge 
\cite{Eto:2004rz,Eto:2006mz,Eto:2006pg} 
(see also \cite{Bruckmann:2007zh,Brendel:2009mp})
\beq
Q_{\rm fractional} = \frac{1}{2\pi} \int d \mu. 
\eeq
We can see that there exist such BPS solutions in the 1d system 
by rewriting the bosonic part of the Euclidean model as
\beq
S_E = \frac{1}{g^2} \int d\tau \, G_{i \bar j} \left( \dot \varphi^i \pm i \xi^i \right) \left( \dot{\bar \varphi}^i \mp i \bar \xi^i \right) \pm \frac{1}{g^2} \int d\mu. 
\eeq
For a given boundary condition, 
this Euclidean action is minimized 
when the following flow equation is satisfied 
(see, e.g. \cite{Dvali:1996bg,Chibisov:1997rc,Isozumi:2004jc,Isozumi:2004va}) 
\beq
\p_\tau \varphi^i = \pm G^{\bar j i} \frac{\p \mu}{\p \bar \varphi^{\bar j}}, 
\eeq
where we have used $\xi^i = i G^{i \bar j} \bar \p_j \mu$. 
This equation describes kink solutions connecting saddle points of $\mu$. 
The on-shell value of the Euclidean action for a kink 
interpolating two saddle points $s$ and $s'$ of $\mu$ is given by
\beq
S_{E,(s,s')} = \left| \frac{\mu_s-\mu_{s'}}{g^2} \right|. 
\eeq
This implies that bion configurations i.e. bound states of kink and antikink 
give typical nonperturbative effects of order $\exp\left(-2S_{E,(s,s')} \right)$. 

\paragraph{Hamiltonian and ground state wave function \\} 
Let us quantize the system by introducing the commutation relation, 
whose nontrivial part is given by
\beq
[\varphi^i , p_j ] = i \delta^i{}_j, \hs{10}
[\bar \varphi^{\bar i}, \bar p_{\bar j}] = i \delta^{\bar i}{}_{\bar j}, \hs{10}
\{\psi^i, \pi_{\psi^j} \} = i \delta^i{}_j, 
\eeq
where $p_i$, $\bar p_{\bar i}$ and $\pi_{\psi^i}$ are the conjugate momenta, 
which can be read off from the Lagrangian as
\beq
p_i = \frac{1}{g^2} \left[ G_{i \bar j} \dot{\bar \varphi}^{\bar j} + i G_{k \bar j} \Gamma^k_{il} \bar \psi^{\bar j} \psi^l \right], \hs{10}
\bar p_{\bar i} = \frac{1}{g^2} G_{j \bar i} \dot \varphi^j, \hs{10}
\pi_{\psi^i} = \frac{i}{g^2} G_{i \bar j} \bar \psi^{\bar j}, 
\eeq
In terms of these canonical coordinates, 
the supercharges can be written as
\beq
Q = \psi^i \left[ p_i + \frac{1}{g^2} G_{j \bar l} \left( \delta^j{}_i \bar \xi^{\bar l} - i \Gamma^j_{ik} \{\psi^k, \bar \psi^{\bar l} \} \right) \right], \hs{10} 
\bar Q = \bar \psi^{\bar i} \left( \bar p_{\bar i} + \frac{1}{g^2} G_{j \bar i} \xi^j \right).
\label{eq:defomedQ}
\eeq
From the superalgebra $H = \{ Q, \bar Q \} - q$, 
we can find the explicit form of the Hamiltonian as
\beq
H &=& g^2 G^{\bar j i} \left( p_i - \frac{1}{g^2} i G_{l \bar m} \bar \psi^{\bar m} \psi^k \Gamma_{ik}^l \right) \bar p_{\bar j} + g^2 \bar \psi^{\bar j} \psi^i R_{i \bar j} + \frac{1}{g^2} G_{i \bar j} \left( \xi^i \bar \xi^{\bar j} - i \nabla_k \xi^i \psi^k \bar \psi^{\bar j} \right) ,
\label{eq:KahlerH2}
\eeq
where $R_{i \bar j}$ is the Ricci curvature $R_{i \bar j} = - \bar \p_{\bar j} \Gamma^k_{ik}$. 

The fermion number operator $F = \frac{1}{g^2} G_{i \bar j} \bar \psi^{\bar j} \psi^i$
commutes with the Hamiltonian, we thus consider the eigenvalue problem of $H$ 
within each sector with a fixed fermion number $F$. 
In the zero fermion sector $F=0$, any state vector satisfies
\beq
\langle \Psi | F |\Psi \rangle = \frac{1}{g^2} \sum_{a=1}^n \big| \big| e^a{}_i \psi^i | \Psi \rangle \big| \big|^2 = 0 
~~~\Longrightarrow~~~ \psi^i |\Psi \rangle = 0,
\label{eq:F0sector}
\eeq
where $e^a{}_i$ denote the vielbein of the target space 
\beq
\sum_{a=1}^n e^a{}_i \overline{e^a{}_j} = G_{i \bar j}. 
\eeq
In this sector, the Hamiltonian reduces to
\beq
H = G^{\bar j i} \left[ g^2 p_i \bar p_{\bar j} + \frac{1}{g^2} \p_i \mu \, \bar \p_{\bar j} \mu - \p_i \bar \p_{\bar j} \mu \right]. 
\label{eq:F0sectorH}
\eeq
We also note that the conserved charge in this sector becomes $q = \xi^i p_i + \bar \xi^{\bar j} \bar p_{\bar j}$. 
The SUSY
ground state, which satisfies $H|\Psi \rangle = 0$, 
is the one annihilated by $\bar p_{\bar i} - \frac{i}{g^2} \bar \p_{\bar i} \mu$, 
and hence the wave function is given by
\beq
\langle \varphi | \Psi \rangle = \exp \left( - \frac{\mu}{g^2} \right).
\eeq


\section{Localization method for $\langle 0|0\rangle$ in K\"ahler quantum mechanics}
\label{app:Loc}
In this appendix, we calculate $\langle 0|0 \rangle$, i.e. the 
generating function for $\langle \mu \rangle$
in K\"ahler QM by means of SUSY localization 
(Duistermaat-Heckman formula).
By introducing the fermionic coordinates, 
$\langle 0|0 \rangle$ can be rewritten as
\beq
\langle 0|0 \rangle = \left( \frac{i}{2} g^2 \right)^n \int d^{2n} \varphi \, d^n \psi \, d^n \bar \psi \, \exp \left( - X \right), 
\hs{10}
X = \frac{2}{g^2} \left( \mu + i G_{i \bar j} \bar \psi^j \psi^i \right). 
\eeq
The integrand is invariant under the following SUSY transformation
\beq
\delta \varphi^i = \psi^i, \hs{10} \delta \psi^i = \xi^i.
\eeq
Note that the square of this SUSY transformation is the holomorphic isometry
$\delta_q \varphi^i = \xi^i,  \delta_q \psi^i = \psi^j \p_j \xi^i$.  
To calculate $\langle 0 | 0 \rangle$, 
let us consider the following integral 
\beq
\left \langle \exp \left( -t \delta V \right) \right \rangle = 
\left( \frac{i}{2} g^2 \right)^n \int d^{2n} \varphi \, d^n \psi \, d^n \bar \psi \, \exp \left( - X - t \delta V \right),
\eeq
where $t$ is a formal parameter and 
$V$ is the following operator which is invariant under the holomorphic isometry $\delta^2 V = \delta_q V = 0$: 
\beq
V =  G_{i \bar j} \xi^i \bar \psi^j + (c.c.). 
\eeq
The invariance under the SUSY transformation implies that 
$\left \langle \exp \left( -t \delta V \right) \right \rangle$
does not depend on $t$, that is, 
\beq
\frac{d}{dt} \left \langle \exp \left( -t \delta V \right) \right \rangle ~=~ -
\left( \frac{i}{2} g^2 \right)^n \int d^{2n} \varphi \, d^n \psi \, d^n \bar \psi \, \delta \Big[ V \exp \left( - X - t \delta V \right) \Big] ~=~ 0.
\eeq
Since $\left \langle \exp \left( -t \delta V \right) \right \rangle$ reduces to $\langle 0 | 0 \rangle$ for $t=0$, 
its $t$-independence implies that 
\beq
\langle 0|0 \rangle = \left \langle \exp \left( -t \delta V \right) \right \rangle ~~~ \mbox{for arbitrary $t$}. 
\eeq
In the $t \rightarrow \infty$ limit, 
the saddle point approximation becomes exact 
and hence the integral can be computed 
only from the data around the saddle point of $\delta V$, 
whose explicit form is given by
\beq
\delta V = G_{i \bar j} \left( \xi^i \bar \xi^{\bar j} + \nabla_k \xi^i \psi^k \bar \psi^{\bar j} \right).
\eeq
This implies that the saddle points are the zeros of $\xi$, i.e. the fixed points of the isometry.
Around each saddle point, $\delta V$ can be expanded by 
rescaling the fluctuation $(\delta \varphi^i, \psi^i) \rightarrow t^{-1} (\delta \varphi^i, \psi^i)$ and 
taking $t \rightarrow \infty$:
\beq
t \delta V = G^s_{i \bar j} \Big[ (M_s \delta \varphi)^i (M_s^\dagger \delta \bar \varphi)^{\bar j} + i (M_s \psi)^i \bar \psi^{\bar j} \Big] + \mathcal O(t^{-1}), 
\eeq
where we have assumed $\xi^i = i (M_s)^i{}_j \delta \varphi^j$ around the saddle point $s$. 
Since $\left \langle \exp \left( -t \delta V \right) \right \rangle$ is independent of $t$, 
we can ignore the subleading terms depending on $t$. 
By performing the Gaussian integration at each saddle point
and collecting the contributions from all the saddle points,
we obtain the generating function $\langle 0 | 0 \rangle$ as
\beq
\langle 0 | 0 \rangle &=& \sum_{s \in \mathfrak S} \left( \frac{i}{2} g^2 \right)^n \int d^{2n} \varphi \, d^n \psi \, d^n \bar \psi \, \exp \left\{ - \frac{2\mu_s}{g^2} - G^s_{i \bar j} \Big[ (M_s \delta \varphi)^i (M_s^\dagger \delta \bar \varphi)^{\bar j} + i (M_s \psi)^i \bar \psi^{\bar j} \Big] \right\} \notag \\
&=& \left(\frac{\pi g^2}{2} \right)^n \sum_{s \in \mathfrak S} \frac{1}{\det M_s} \exp \left( - \frac{2\mu_s}{g^2} \right).
\eeq


\section{Perturbative part of $E^{(2)}$ in $\C P^{N-1}$ quantum mechanics}\label{appendix:E2CPN}
In this appendix, we calculate the perturbative part of $E^{(2)}$ in the $\C P^{N-1}$ model. 
The leading order correction to the wave function can be obtained 
by solving the Schr\"odinger equation expanded around $\epsilon=1$. 
Its asymptotic behavior in the weak coupling limit is given by
\beq
\Psi^{(1)} = \frac{N}{2} e^{-\frac{\mu}{g^2}} \log \frac{1}{1+\sum_{k=1}^{N-1}|\varphi^k|^2} + \cdots, 
\eeq
where $\cdots$ denotes nonperturbative corrections. 
From the relation $E^{(2)} = - \langle \Psi^{(1)}| H_{\epsilon=1} | \Psi^{(1)}\rangle /\langle 0 | 0 \rangle$, 
we find that
\beq
E^{(2)} = -\frac{N^2 \pi^{N-1} g^{2N}}{4\langle 0 | 0 \rangle} \int_{\Delta_{N-1}} dx_1 \cdots dx_{N-1} \, \frac{g^2\sum_{i=1}^{N-1} x_i}{1-g^2\sum_{i=1}^{N-1}x_i} \, \exp \left( -2 \sum_{i=1}^{N-1} m_i x_i \right)  + \cdots, 
\label{eq:E2_1_CPN}
\eeq
where we have changed the integration variables from
$|\varphi^i|$ to $x_i = \frac{1}{g^2} \frac{|\varphi^i|^2}{1+ \sum_{k=1}^{N-1}|\varphi^k|^2}$.
The integration domain is the $(N-1)$-simplex
\beq
\Delta_{N-1} = \left\{ (x_1,\cdots,x_{N-1}) \in \R^{N-1} \ \Big| \ x_i > 0 , \, x_1 + \cdots + x_{N-1} \leq \frac{1}{g^2} \right\}.
\label{eq:domainCPN}
\eeq
Expanding the integrand with respect to $g$, we obtain
\beq
E^{(2)} = -\frac{N^2 \pi^{N-1} g^{2N}}{4\langle 0 | 0 \rangle} \sum_{n=1}^\infty \left( - \frac{g^2}{2} \sum_{i=1}^{N-1} \frac{\p}{\p m_i} \right)^n \int_{\Delta_{N-1}} dx_1 \cdots dx_{N-1} \, \exp \left( -2 \sum_{i=1}^{N-1} m_i x_i \right) + \cdots .
\nonumber\\
\label{eq:E2_2_CPN}
\eeq
Since the condition $x_1 + \cdots + x_{N-1} \leq 1/g^2$ can be ignored in the weak coupling limit, 
the perturbation series can be obtained 
by integrating over the region with $x_i>0$. 
Then we obtain the following perturbation series
\beq
E^{(2)}_0 =  - \frac{N^2}{4} g^2 \left( \prod_{i=1}^{N-1} \frac{g^2}{m_i} \right)^{-1} \sum_{n=1}^\infty \left( - \frac{g^2}{2} \sum_{i=1}^{N-1} \frac{\p}{\p m_i} \right)^n\left( \prod_{i=1}^{N-1} \frac{g^2}{m_i} \right),
\label{eq:E20_CPN}
\eeq
where we have used $\langle 0 | 0 \rangle = \prod_{i=1}^{N-1} \frac{\pi g^2}{2m_i} + \cdots$. 
This is a divergent series since the coefficient of $g^n$ is of order $n!$. 
By the Borel resummation, $E^{(2)}_0$ can be rewritten as
\beq
E^{(2)}_0 
&=& \frac{N^2}{4} g^2 \left( \prod_{i=1}^{N-1} \frac{g^2}{m_i} \right)^{-1} \int_0^\infty dt \, e^{-t} \left[ \prod_{i=1}^{N-1} \frac{g^2}{m_i} - \prod_{i=1}^{N-1} \frac{g^2}{m_i-\frac{tg^2}{2}} \right],
\label{eq:BSE20_CPN}
\eeq
where we have used
$\sum_{n=1}^\infty \frac{1}{n!} \left( - \frac{t g^2}{2} \sum_{i=1}^{N-1} \frac{\p}{\p m_i} \right)^n f(m_i)  = f( m_i - \textstyle \frac{tg^2}{2}) - f(m_i)$.
Using the partial fraction decomposition
\beq
\prod_{i=1}^{N-1} \frac{1}{m_i-\frac{tg^2}{2}} &=& \sum_{j=1}^{N-1} \frac{1}{m_j-\frac{tg^2}{2}} \prod_{i=1,\,i\not=j}^{N-1} \frac{1}{m_j-m_i}, 
\eeq
we obtain the perturbative part of the second order expansion coefficient as 
\beq
E^{(2)}_0 &=& \frac{N^2}{4} \left[ g^2 + \sum_{i=1}^{N-1} 2 m_i A_i \int_0^\infty dt \frac{e^{-t}}{t-\frac{2m_i}{g^2 \pm i 0}} \right].
\eeq


\section{Quasi moduli space}
\label{appendix:quasi_moduli}
When a saddle point has a nearly flat direction, 
which corresponds to an eigenmode 
whose mass (frequency) vanishes 
in the weak coupling limit $g \rightarrow 0$, 
the Gaussian approximation is not applicable 
for evaluating the contribution of that saddle point to the path integral.  
In such a case, 
we need to integrate all the way along the nearly flat directions, 
which are parameterized by the quasi-moduli parameters 
$\eta^\alpha$. 
Let $\varphi^i_\eta$ be the configuration 
along the nearly flat direction,  
which we define as the configuration such that 
$\frac{\p \varphi_\eta^i}{\p \eta^\alpha}$ is 
proportional to the quasi-zero modes 
at the saddle point and 
the equation of motion is satisfied 
up to terms proportional to $\frac{\p \varphi^i_\eta}{\p \eta^\alpha}$: 
\beq
\frac{\delta S \,}{\delta \varphi^i} \bigg|_{\varphi^i =\varphi^i_\eta} = ~ \sum_{\alpha} c_\alpha(\eta) \frac{\p \varphi_\eta^i}{\p \eta^\alpha}. 
\eeq
By using this ansatz on the quasi-moduli space, 
we decompose the field as 
\beq
\varphi^i = \varphi_\eta^i + g \delta \varphi^i_{\perp}, 
\eeq
where $\delta \varphi^i_{\perp}$ denotes 
the fluctuation containing all the massive modes, 
which are orthogonal to the quasi-zero modes 
$\frac{\p \varphi_s^i}{\p \eta^\alpha}$.
Then the action can be schematically expanded as
\beq
S[\varphi] = S[\varphi_\eta] + g^2 \delta\bar{\varphi}^i \Delta_{\bar ij} \delta \varphi^{j} + \mathcal O(g^4),
\eeq
where $\Delta_{ij}$ is the differential operator appearing 
in the linearized equation of motion. 
The absence of the linear term indicates that 
the quasi-moduli parameterize the bottom of the valley of the action. 
In the weak coupling limit, the path integral for the partition function reduces to the quasi-moduli integral
\beq
Z \approx \int dv \frac{1}{\det \Delta_{\bar ij}} \, \exp \left( - \frac{V_{\rm eff}}{g^2} \right),
\eeq
where $dv$ is the volume form of the quasi-moduli space, 
$\det \Delta$ is the one-loop determinant
and $V_{\rm eff}$ is the kink-antikink effective potential 
which can be obtained by substituting the kink-antikink ansatz into the original action: 
\beq
V_{\rm eff}(\eta) = S[\varphi_\eta]. 
\eeq

\section{Kink-antikink effective potential}
\label{appendix:effpot}
In this appendix, we derive a general formula for the effective potential between a pair of well-separated kink and antikink. 
To find the effective potential, 
it is necessary to find an appropriate ansatz 
for the kink-antikink configuration $\varphi_\eta^i$ parametrized by the quasi-moduli parameters $\eta^\alpha$. 
Since the kink and antikink are well-separated 
in the bion configuration for small $g$, 
their positions $\tau_k$ and $\tau_{\bar k}$ become 
almost free parameters 
and hence can be interpreted as the quasi-moduli parameters 
($\eta^\alpha = \tau_k$, $\tau_{\bar k}$ and other possible internal degrees of freedom). 
 
When the kink and antikink are well-separated so that 
$|\tau_k - \tau_{\bar k}|$ is much larger 
than any length scale in the model, 
then the kink-antikink configuration can be schematically written as
\beq
\varphi^i_\eta = 
\left\{ 
\begin{array}{ll} 
\varphi_k^i + \delta \varphi_{\bar k}^i  + \cdots & \tau \approx \tau_k \\
v^i + \delta \varphi_k^i + \delta \varphi_{\bar k}^i + \cdots & \tau \approx \tau_0 \\
\varphi_{\bar k}^i + \delta \varphi_k^i + \cdots & \tau \approx \tau_{\bar k} 
\end{array}
\right.,
\label{eq:kink_antikink_ansatz}
\eeq
where $v^i$ are the vacuum expectation values and 
$\tau_0$ is a point in between the kink and antikink 
$(\tau_k \ll \tau_0 \ll \tau_{\bar k})$ 
such that the tail from the kink (antikink) can be approximated 
by a small perturbation 
$\delta \varphi_k^i$ ($\delta \varphi_{\bar k}^i$) 
for $\tau > \tau_0$ ($\tau < \tau_0$). 

Let us decompose $V_{\rm eff}$ into the contributions 
from $\tau > \tau_0$ and $\tau < \tau_0$:
\beq
V_{\rm eff} = S[\varphi_{k \bar k}^i] = S_{\tau>\tau_0} + S_{\tau<\tau_0}
\eeq
Since the kink contribution can be treated 
as the small perturbation $\delta \varphi_k^i$ 
in the region $\tau>\tau_0$, 
$S_{\tau<\tau_0}$ is approximately given by
\beq
S_{\tau>\tau_0} &=& S[\varphi_{\bar k}^i] + \int d \tau \, \left( \delta \varphi_k^i \frac{\p \mathcal L}{\p \varphi^i} + \delta \dot \varphi_k^i \frac{\p \mathcal L}{\p \dot \varphi^i} \right) \bigg|_{\varphi^i = \varphi^i_{\bar k}} + \cdots \notag \\
&=& S[\varphi_{\bar k}^i] - \left( \delta \varphi_k^i \frac{\p \mathcal L}{\p \dot \varphi^i} \bigg|_{\varphi^i = \varphi^i_{\bar k}} \right)_{\tau=\tau_0} + \cdots,
\eeq
where we have used the fact that $\varphi_k^i$ satisfies 
the Euler-Lagrange equation and 
the contribution from $\tau = \infty$ is trivial. 
In the similar way, we can calculate the contribution from the 
region $\tau<\tau_0$. Adding two contributions, we obtain
\beq
V_{\rm eff} = S[\varphi_k^i] + S[\varphi_{\bar k}^i] + \left( \delta \varphi_{\bar k}^i \frac{\p \mathcal L}{\p \dot \varphi^i} \bigg|_{\varphi^i = \varphi^i_k} - \delta \varphi_k^i \frac{\p \mathcal L}{\p \dot \varphi^i} \bigg|_{\varphi^i = \varphi^i_{\bar k}} \right)_{\tau=\tau_0} + \cdots. 
\label{eq:Veff_formula}
\eeq

Let $\delta \varphi^i$ be the fluctuations around the VEVs $v^i$ 
orthonormalized so that the expanded Lagrangian takes the form
\beq
\mathcal L = \frac{1}{2} (\delta \dot \varphi^i)^2 
+ \frac{m_i^2}{2} (\delta \varphi^i)^2 + \cdots.
\eeq
Then, for $\tau \approx \tau_0$, 
the small deviations $\delta \varphi_k^i$ 
and $\delta \varphi_{\bar k}^i$ 
in \eqref{eq:kink_antikink_ansatz} take the following forms: 
\beq
\delta \varphi_k^i = {\mathcal A}^i e^{-m_i \tau}, \hs{10} 
\delta \varphi_{\bar k}^i = {\mathcal B}^i e^{m_i \tau},
\label{eq:deviations}
\eeq
where ${\mathcal A}^i$ and ${\mathcal B}^i$ are 
constants depending on the quasi-moduli parameters. 
Substituting into \eqref{eq:Veff_formula}, 
we obtain the effective potential as
\beq
V_{\rm eff} \approx S[\varphi_k^i] + S[\varphi_{\bar k}^i] 
+ \sum_i 2 m_i {\mathcal A}^i {\mathcal B}^i. 
\eeq
Therefore, the effective kink-antikink potential can be obtained 
by determining the coefficients 
${\mathcal A}^i$ and ${\mathcal B}^i$ in \eqref{eq:deviations}. 

In the presence of a potential term induced by the fermion projection, 
the vacuum $\varphi^i = v_i$ between kink and antikink 
may not be the true minimum but can be lifted, 
so that the following confining potential term is 
induced in the effective potential 
\beq
V_{\rm eff} ~\approx~ S[\varphi_k^i] + S[\varphi_{\bar k}^i] 
+ \sum_i 2 m_i {\mathcal A}^i {\mathcal B}^i + |\tau_k - \tau_{\bar k}| \delta S,
\eeq
where $\delta S$ is the difference of the action between 
$\varphi^i = v^i$ and the true minimum:
\beq
\delta S ~\equiv~ S[\varphi^i = v^i] - S[\varphi^i=\mbox{true minimum}] ~>~ 0. 
\eeq
All the effective kink-antikink potential used 
in this paper can be obtained 
by substituting the explicit forms of ${\mathcal A}^i$, ${\mathcal B}^i$ and $\delta S$ in each model. 


\section{Lefschetz thimble analysis of quasi-moduli integral}
\label{appendix:QMI}
In this appendix, we calculate the following quasi-moduli integral of the form \eqref{eq:Ii_SG}: 
\beq
I \ = \ \int_{\mathcal C} dy \, \exp \left[ - V(y) \right], \hs{10}
V(y) \equiv a e^{-y} + b y, \hs{5} {\rm Re} \, b > 0,
\eeq
where $\mathcal C$ denotes the integration contour along the real axis.
The parameter $a$ is positive for kinks of the same type and negative for a kink-antikink pair. 
Note that the divergence of $V$ for ${\rm Re} \, y \rightarrow -\infty$ is an artifact of the approximation
and the form of $V$ around its saddle points is relevant 
for the asymptotic behavior of the integral for small $g$ (large $a$). 

\begin{figure}[h]
\includegraphics[width=80mm]{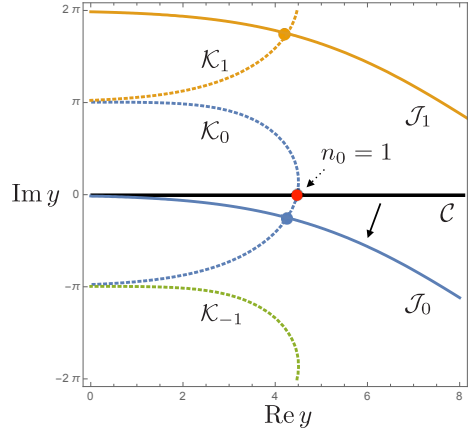}
\caption{The Lefschetz thimbles $\mathcal J_q$ and their duals $\mathcal K_q$.}
\label{fig:theta_0}
\end{figure}

\begin{figure}[h]
\begin{minipage}{0.45\hsize}
\centering
\includegraphics[width=80mm]{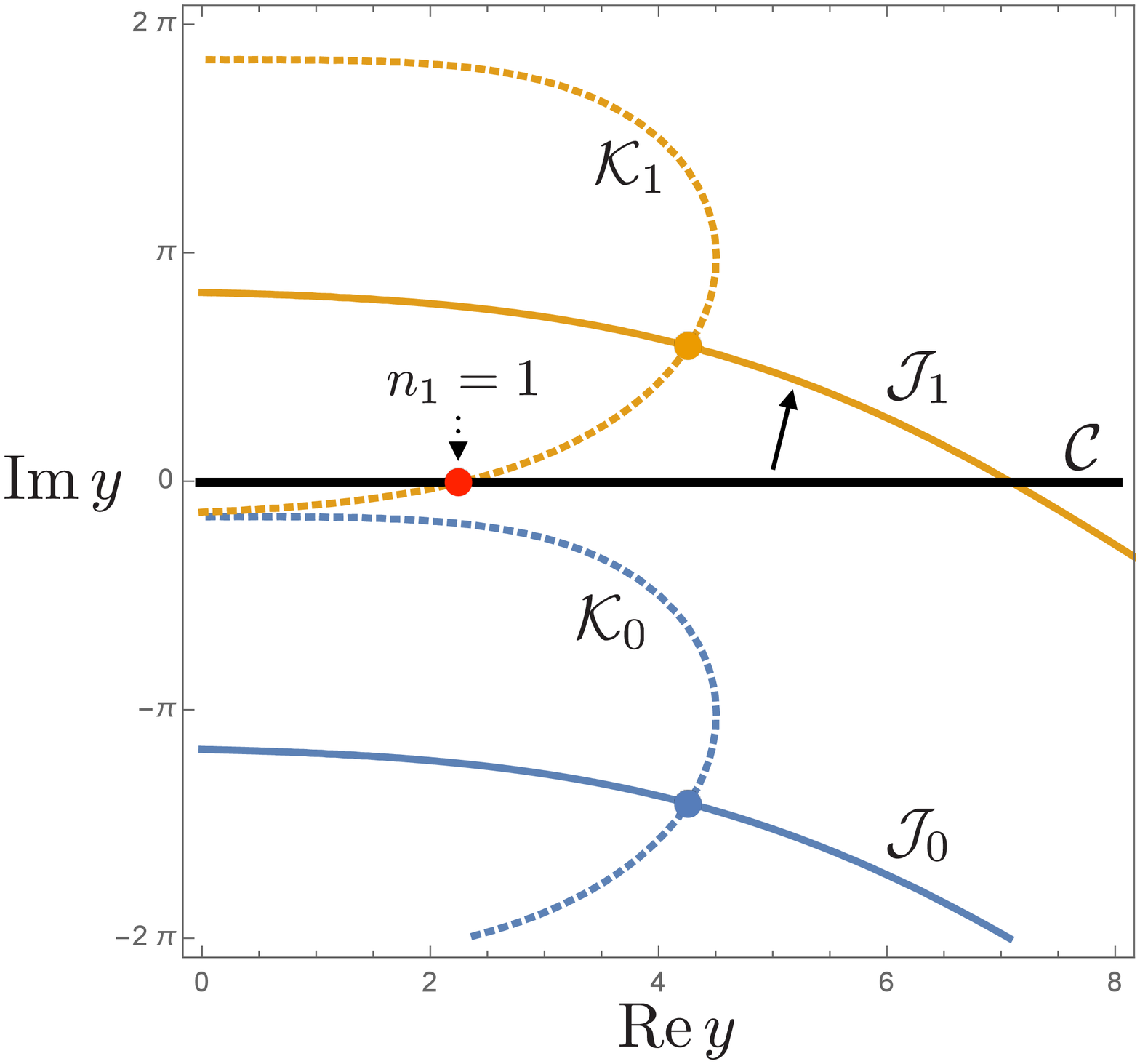} \\
\hs{7} (b) ~$\theta > 0$
\end{minipage}
\begin{minipage}{0.45\hsize}
\centering
\includegraphics[width=80mm]{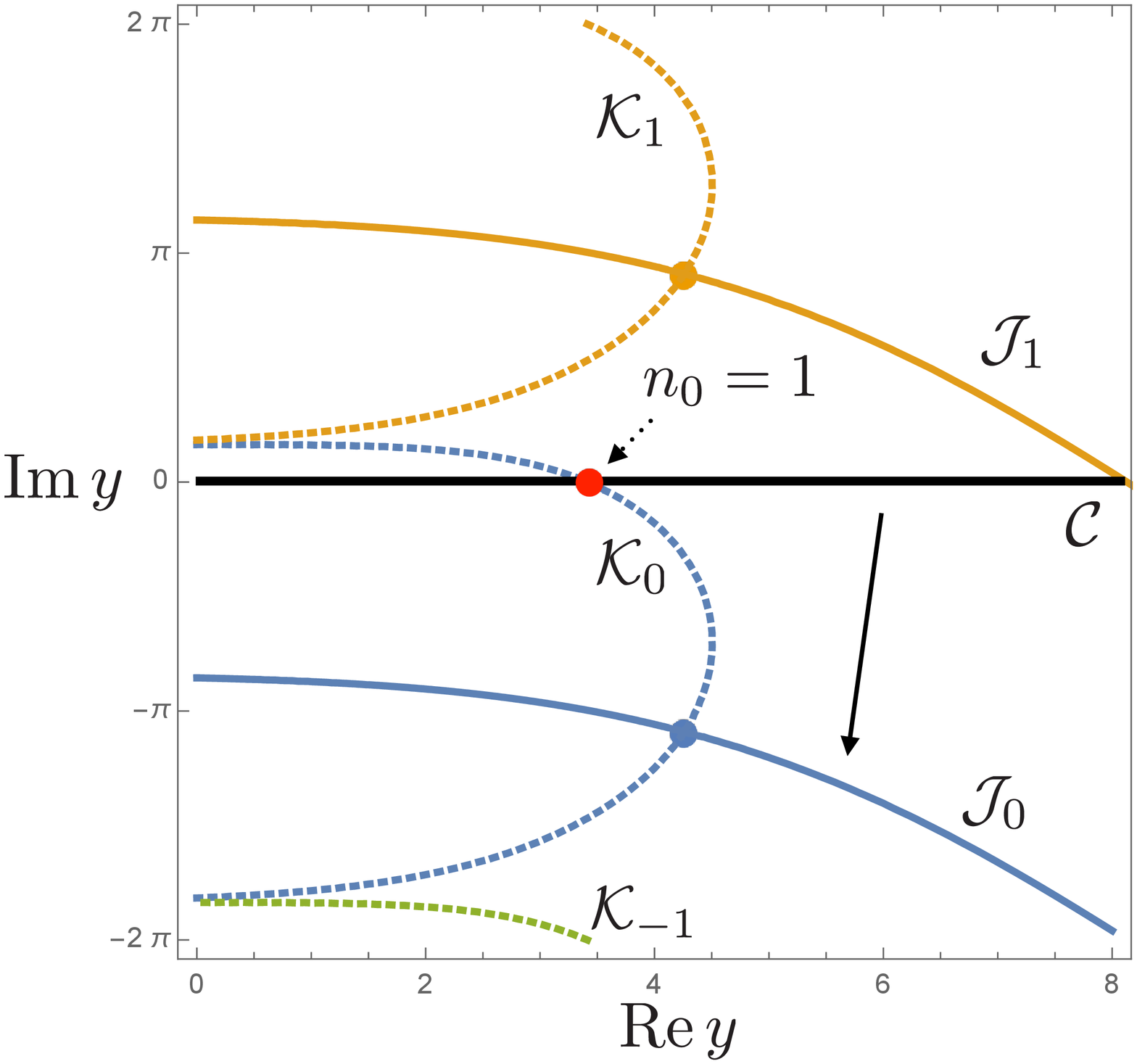} \\
\hs{7} (b) ~$\theta < 0$
\end{minipage}
\caption{Stokes phenomenon at $\arg a = - \pi$ ($\theta = - \pi - \arg a = 0$). 
The original integration contour $\mathcal C$ intersects 
with $\mathcal K_1$ ($\mathcal K_0$) for $\theta > 0$ ($\theta < 0$) 
and hence $\mathcal C$ is deformed to $\mathcal J_1$ ($\mathcal J_0$).}
\label{fig:stokes}
\end{figure}

The flow equation, which determines the Lefschetz thimbles, is given by 
\beq
\frac{\p y}{\p t} = \overline{\frac{\p V}{\p y}} = - \bar a e^{-\bar y} + \bar b.
\eeq
The Lefschetz thimbles $\mathcal J_q$ and their duals $\mathcal K_q$ 
for the saddle points 
\beq
y = \log \frac{a}{b} + 2 \pi i q, \hs{10} q \in \Z,
\eeq 
are depicted in Fig.\,\ref{fig:theta_0} ($a > 0$) and Fig.\,\ref{fig:stokes} ($a < 0$). 
For $a > 0$, 
the real axis $\mathcal C$ intersects 
with the dual thimble of the saddle point $q=0$, 
so that the integral is given by
\beq
I \ = \ \int_{\mathcal J_0} dy \, \exp \left[ - V(y) \right] \ = \ a^{-b} \Gamma(b).
\eeq

For $a<0$, 
the Stokes phenomena occurs 
when we vary $\arg a$ around $\arg a = -\pi$ as shown in Fig.\,\ref{fig:stokes}. 
Thus we obtain the following ambiguous result for $a < 0$: 
\beq
I \ = \ 
\left\{ 
\begin{array}{c} 
\int_{\mathcal J_1} dy \, \exp \left[ - V(y) \right] \\ 
\int_{\mathcal J_0} dy \, \exp \left[ - V(y) \right] 
\end{array} 
\right. \ 
= \ 
|a|^{-b} \exp( \mp \pi i b) \, \Gamma(b) & \mbox{~~for $\theta \equiv - \pi - \arg a = \pm 0$}.
\eeq
Combining these result, we obtain Eq.\,\eqref{eq:Ii2_SG}. 

\section{One-loop determinant in $\C P^{N-1}$ quantum mechanics}\label{appendix:one-loop}
In this appendix, we calculate the one-loop determinant 
around the $i$-th single bion background $\varphi^i_\eta$ in the $\C P^{N-1}$ model. 
It is convenient to normalize the fluctuations in the background of the $i$-th bion as
\beq
\varphi^i = \varphi_\eta^i + g (1+|\varphi_\eta^i|^2) \, \delta \varphi^i, \hs{10} 
\varphi^j = g \sqrt{1+|\varphi_\eta^i|^2} \, \delta \varphi^j ~~~ (j \not = i). 
\eeq
It was shown in \cite{Fujimori:2016ljw} that the contribution from $\delta \varphi^i$ to the one-loop determinant 
gives the overall factor $8m_i^4/\pi g^2$ in Eq.\,\eqref{eq:QMI_CPN} . 
To calculate the contribution from $\delta \varphi^j$, 
let us consider the linearized equation for $\delta \varphi^j$ 
in the background $\varphi^i_\eta$. 
\beq
\Delta_j \delta \varphi^j \ = \ ( -\p_\tau^2 + {\mathcal V}_{i,j} ) \delta \varphi^j \ = \ 0, 
\label{eq:Lineq_CPN}
\eeq
where $\Delta_j = -\p_\tau^2 + {\mathcal V}_{i,j}$ is 
the differential operator appearing in Eq.\,(\ref{eq:SVeff_CPN}). 
The determinant $\Delta_j$ can 
be read off from the following asymptotic behavior of 
the solution of the linearized equation 
(see e.g. Appendix B of \cite{Fujimori:2016ljw})
\beq
1 = \lim_{\tau \rightarrow - \infty} e^{-m_j \tau} \delta \varphi^j, \hs{10} 
\det \Delta_j = \lim_{\tau \rightarrow \infty} e^{-m_j \tau} 
\delta \varphi^j.
\label{eq:Det_CPN}
\eeq
In the weak coupling limit $g \rightarrow 0$, the single bion 
can be viewed as a pair of well-separated kink and antikink, 
for which $ {\mathcal V}_{i,j}$ can be approximated as
\beq
 {\mathcal V}_{i,j} \approx \left\{ 
\begin{array}{ll} 
m_j^2 & \tau \ll \tau_- \\ 
\mathcal U_{i,j}(\tau - \tau_-)  & \tau \sim \tau_- \\ 
(m_j-m_i)^2 & \tau_- \ll \tau \ll \tau_+ \\ 
\bar{\mathcal U}_{i,j}(\tau - \tau_+) & \tau \sim \tau_+ \\
m_j^2 & \tau_+ \ll \tau
\end{array} \right.,
\label{Vij_CPN}
\eeq
where $\mathcal U_{i,j}$ and $\bar{\mathcal U}_{i,j}$ 
are the following background potential felt by $\delta \varphi^j$
due to the BPS kink $\varphi^i_{k} = e^{m_i (\tau-\tau_-)}$ and 
the anti-BPS kink $\varphi^i_{\bar k} = e^{-m_i (\tau-\tau_+)}$: 
\beq
\mathcal U_{i,j}(\tau-\tau_-) &=& \left( m_j - \frac{m_i}{1+e^{-2m_i(\tau-\tau_-)}} \right)^2 + \p_\tau \left( m_j - \frac{m_i}{1+e^{-2m_i(\tau-\tau_-)}} \right), \notag \\
\bar{\mathcal U}_{i,j}(\tau-\tau_+) &=& \ \left( m_j - \frac{m_i}{1+e^{2m_i(\tau-\tau_-)}} \right)^2 - \ \p_\tau \left( m_j - \frac{m_i}{1+e^{2m_i(\tau-\tau_-)}} \right). 
\eeq
In each region, the linearized equation \eqref{eq:Lineq_CPN} 
can be solved as
\beq
\delta \varphi^j \approx \left\{ 
\begin{array}{ll} 
e^{m_j \tau} & \tau \ll \tau_- \\
f_- & \tau \sim \tau_-  \\
a'_j \, e^{(m_j-m_i) \tau} + b'_j \, e^{-(m_j-m_i) \tau} 
& \tau_- \ll \tau \ll \tau_+ \\ 
f_+ & \tau \sim \tau_+ \\ 
{\rm det}\Delta_j \, e^{m_j \tau} + b''_j \, e^{-m_j \tau} & \tau_+ \ll \tau
\end{array} \right..
\label{eq:xii_CPN}
\eeq
with 
\beq
f_- &=& \frac{c_j \, e^{m_j \tau}}{\sqrt{1+e^{2m_i(\tau-\tau_-)}}} 
+ \frac{d_j \, e^{-m_j \tau}}{\sqrt{1+e^{2m_i(\tau-\tau_-)}}} 
\left( 1 + \frac{m_j}{m_j-m_i} e^{2m_i(\tau-\tau_-)} \right), \\
f_+ &=& \frac{c'_j \, e^{-m_j \tau}}{\sqrt{1+e^{-2m_i(\tau-\tau_+)}}} 
+ \frac{d'_j \, e^{m_j \tau}}{\sqrt{1+e^{-2m_i(\tau-\tau_+)}}} 
\left( 1 + \frac{m_j}{m_j-m_i} e^{-2m_i(\tau-\tau_+)} \right). 
\eeq
Connecting these solutions, 
we find that the coefficients are related as 
\beq
c_j = 1, \hs{10} a'_j = e^{m_j \tau_-} , \hs{10} 
d'_j = \frac{m_j-m_i}{m_j} e^{m_i (\tau_--\tau_+)} = \det \Delta_j.
\label{eq:cad_CPN}
\eeq
Therefore, the one-loop determinant is given by
\beq
\prod_{j=1,\,j \not = i}^{N-1} \det \Delta_j^{-1} 
= A_i \, e^{(N-2)m_i \tau_r}, 
\label{eq:prodDet_CPN}
\eeq 
where $\tau_r=\tau_+-\tau_-$ is the relative position 
and the constant $A_i$ is defined in (\ref{eq:constant-Ai}).


\end{document}